\documentclass[a4paper,fleqn]{cas-sc}

\usepackage[numbers,sort&compress]{natbib}
\usepackage{csquotes}
\usepackage{xcolor}
\usepackage[colorinlistoftodos,prependcaption]{todonotes}
\usepackage[inline]{enumitem}
\usepackage{xspace}
\usepackage{subcaption}
\usepackage{tablefootnote}
\usepackage{multicol}
\usepackage{longtable}
\usepackage[nohyperlinks,nolist]{acronym}
\usepackage{wasysym} 
\usepackage{float}
\usepackage{amssymb}
\usepackage{pifont}
\newcommand{\cmark}{\ding{51}}%
\newcommand{\xmark}{\text{\ding{55}}}

\def\tsc#1{\csdef{#1}{\textsc{\lowercase{#1}}\xspace}}
\tsc{WGM}
\tsc{QE}
\tsc{EP}
\tsc{PMS}
\tsc{BEC}
\tsc{DE}

\usepackage{placeins}

\begin{document}

\shorttitle{Portable Medical Imaging in Modern Healthcare: Fundamentals, AI-Based Taxonomy, Image Quality, and Open Challenges}

\shortauthors{Y. Habchi et al.}
                      
\title [mode = title]{Portable Medical Imaging in Modern Healthcare: Fundamentals, AI-Based Taxonomy, Image Quality, and Open Challenges}

\vskip2mm

\author[1]{Yassine Habchi}[orcid=0000-0002-8764-9675]
\ead{habchi@cuniv-naama.dz}
\cormark[1]
\credit{Conceptualization; Methodology; Resources; Investigation; Writing original draft; Writing, review, and editing}

\author[2]{Hamza Kheddar}[orcid=0000-0002-9532-2453]
\ead{kheddar.hamza@univ-medea.dz}
\credit{Conceptualization; Methodology; Data Curation; Resources; Investigation; Visualization;  Writing original draft; Supervising; Writing, review, and editing}

\author[3]{Muhammad Ali Qureshi}[orcid=0000-0003-4390-2461]
\ead{ali.qureshi@iub.edu.pk}
\credit{Conceptualization; Investigation; Visualization; Writing, review, and editing}

\author[4]{Mohamed Seghier}[orcid=0000-0002-1146-8800]
\ead{mohamed.seghier@ku.ac.ae}
\credit{Conceptualization; Investigation; Visualization;  Writing, review, and editing}

\author[5]{Azeddine Beghdadi}[orcid=0000-0002-5595-0615]
\ead{beghdadi@sorbonne-paris-nord.fr}
\credit{Conceptualization; Investigation; Visualization;  Supervising; Project management; Writing, review, and editing}

\address[1]{Faculty of Technology, University Salhi Ahmed, Naama, Algeria}
\address[2]{LSEA Laboratory, Department of Electrical Engineering, University of Medea, 26000, Algeria}
\address[3]{Department of Information \& Communication Engineering, The Islamia University of Bahawalpur, Pakistan}
\address[4]{Department of Biomedical Engineering and Biotechnology, Khalifa University of Science and Technology, Abu Dhabi, UAE}

\address[5]{Sorbonne Paris Nord University Villetaneuse, France}

\tnotetext[1]{Y. Habchi is the corresponding author.}

\begin{abstract}
Portable medical imaging (PMI) has emerged as an important solution for point-of-care diagnosis in emergency, rural, and resource-limited settings where conventional imaging infrastructure is not readily available. Modalities such as portable computed tomography, portable magnetic resonance imaging, portable ultrasound, and wireless capsule endoscopy improve access to timely diagnosis, but they remain highly vulnerable to image-quality degradation caused by motion artifacts, environmental interference, hardware limitations, and unstable acquisition conditions. This review provides a systematic and quality-centered synthesis of recent advances in PMI. It introduces a taxonomy of AI-based PMI methods spanning machine learning, deep learning, transfer learning, and Transformer-based approaches, and examines their roles in image enhancement, reconstruction, quality assessment, detection, and classification. The review also analyzes PMI devices, sensing pipelines, modality-specific distortions, evaluation metrics, and publicly available datasets. In contrast to existing surveys that are mainly modality-driven or application-focused, this work emphasizes the relationship between image quality, AI robustness, and clinical usability in portable settings. Finally, it identifies current research gaps and outlines future directions toward reliable, interpretable, and clinically deployable PMI systems.
\end{abstract}


\maketitle
\FloatBarrier
\section{Introduction} \label{sec1}

Remote areas face significant challenges in accessing healthcare due to geographical  isolation and limited medical resources. These challenges intensify during natural disasters such as earthquakes and floods, when infrastructure damage hinders the arrival of emergency teams and medical supplies. Similar constraints also affect veterinary care, where treating large animals—such as horses, camels, and cattle—may require urgent field surgeries with specialized equipment inside well-equipped clinics. Transporting heavy medical equipment without suitable means of transportation further delays treatment, increasing the risk of complications and fatalities. During epidemics, these regions experience additional pressure as patient numbers exceed healthcare capacity, weakening the system’s ability to respond effectively \cite{kavanagh2023scoping,mazari2023deep,ali2023review}. To overcome these obstacles, innovative solutions have been developed, including telemedicine for remote diagnosis, mobile medical units for on-site care, and \ac{PMI} technologies that allow imaging directly at the patient’s location. These approaches minimize patient transport and maintain continuity of care when infrastructure is limited. \Ac{PMI} represents a major advancement in healthcare delivery, providing diagnostic imaging capabilities in settings that lack hospital infrastructure. It enables clinicians to obtain high-quality images rapidly, facilitating timely medical decisions \cite{bian2024review, altaf2023utilization, ploger2024evaluating}. Major \ac{PMI} modalities include \ac{PCT} \cite{wang2024deep}, \ac{PUS} \cite{perez2024comparison}, \ac{PMRI} \cite{groteklaes2025case}, and \Ac{WCE} \cite{cao2024robotic}. Their compact design and portability make them particularly valuable in rural and resource-limited environments.

\begin{figure}
\centering
\includegraphics[width=10cm, height=10cm]{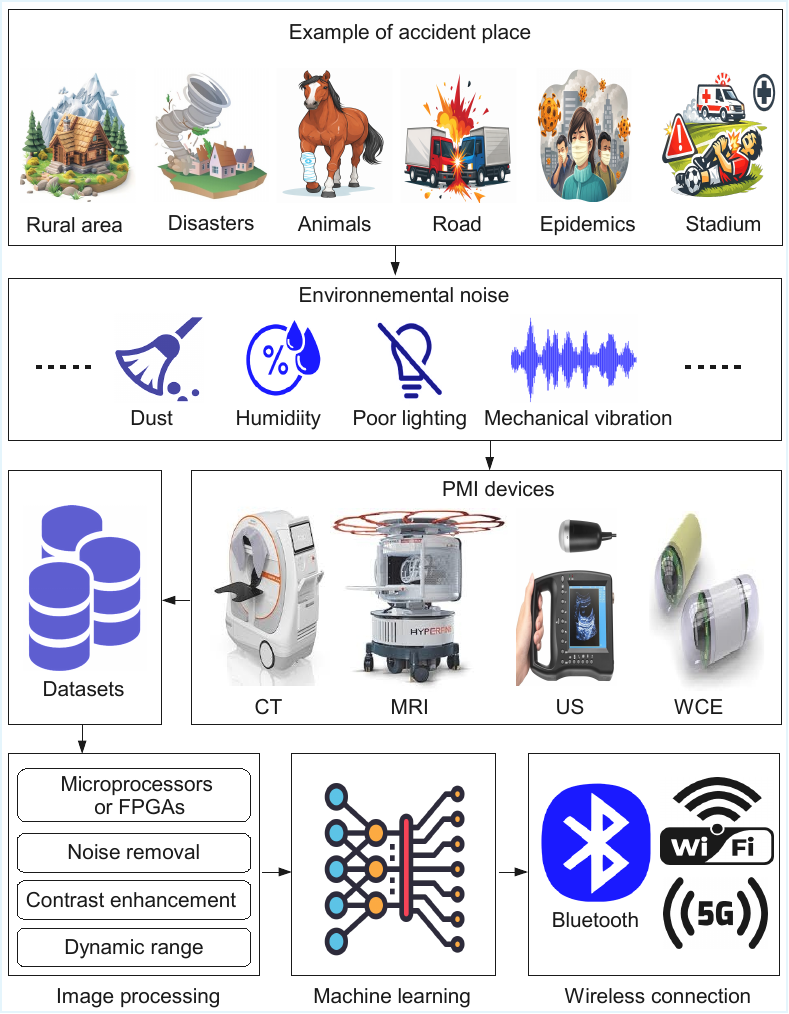}
\caption{An example block diagram of PMI fundamentals. }
\label{figexampleofblockdiagramofPMI}
\end{figure}


\begin{acronym}[AWGN]  
\acro{AI}{artificial intelligence}
\acro{AUC}{area under the curve}
\acro{Acc}{accuracy}

\acro{CT}{computed tomography}
\acro{CNN}{convolutional neural network}
\acro{CAD}{computer-aided detection}

\acro{DL}{deep learning}
\acro{DICE}{dice similarity coefficient}

\acro{FDCNN}{feature-guided denoising convolutional neural network}
\acro{FR}{full reference}
\acro{F1}{f1-score}
\acro{FPGA}{field programmable gate array}
\acro{FSIM}{feature similarity index}

\acro{IQA}{image quality assessment}
\acro{IoT}{internet of things}
\acro{IoU}{intersection over union}

\acro{GAN}{generative adversarial network}

\acro{LLM}{large language model}

\acro{MWHI}{microwave head imaging system}
\acro{ML}{machine learning}
\acro{MRI}{magnetic resonance imaging}
\acro{MIQA}{medical image quality assessment}
\acro{MIQ}{medical image quality}
\acro{MSE}{mean squared error}
\acro{mAP}{mean average precision}
\acro{MAE}{mean absolute error}

\acro{NR}{no reference}
\acro{NQM}{noise quality measure}
\acro{NCC}{normalized cross-correlation}

\acro{PMI}{portable medical imaging}
\acro{POCUS}{portable point-of-care ultrasound}
\acro{PSNR}{peak signal-to-noise ratio}
\acro{PCT}{portable CT}
\acro{PMRI}{portable MRI}
\acro{PUS}{portable US}
\acro{Pre}{precision}
\acro{ADC}{analog-to-digital converter}
\acro{RMW}{reconstructe microwave}
\acro{DRL}{deep reinforcement learning}
\acro{RR}{reduced reference}
\acro{RMSE}{root mean square error}


\acro{TL}{transfer learning}

\acro{SSIM}{structural similarity index}
\acro{SVM}{support vector machine}
\acro{Sen}{sensivity}
\acro{Spe}{specificity}
\acro{SNR}{signal-to-noise ratio}
\acro{IMU}{inertial measurement unit}
\acro{ViT}{vision transformer}
\acro{VIF}{visual information fidelity}

\acro{US}{ultrasound}

\acro{WCE}{wireless capsule endoscopy}
\end{acronym}


Figure \ref{figexampleofblockdiagramofPMI} illustrates the essential workflow and components of \ac{PMI} devices. On the above, modern portable imaging modalities are shown, including \ac{PCT}, \ac{PMRI}, and \ac{PUS}. These rely on advanced sensors to capture diagnostic data in diverse clinical environments. Subsequently, the data are handled by processing units such as microprocessors or \acp{FPGA} that employ techniques like noise reduction, contrast enhancement, and dynamic range optimization to enhance image clarity. Next, \ac{ML} algorithms analyze and interpret the images, supporting faster and more accurate diagnoses. Finally, the processed results are transmitted via wireless communication technologies (Bluetooth, Wi-Fi, 5G), enabling remote access, telemedicine, and cloud-based healthcare solutions. The diagnostic accuracy of \ac{PMI} depends strongly on image quality. However, several factors can degrade performance: environmental noise (e.g., dust, humidity, and poor lighting), mechanical vibration during field operation, and limited sensor resolution compared to stationary hospital systems. Technical issues such as miscalibration, long-distance transport, or inadequate maintenance may further impair sensitivity. Moreover, insufficient operator training can lead to suboptimal device use, reducing image clarity and interpretability \cite{edwards2023assessing,salimi2022ultrasound, wang2019high,liang202214}.  These limitations highlight the need either for reliable image-quality assurance mechanisms tailored to \ac{PMI} conditions or a robust \ac{ML} algorithm capable of extracting insight from low image-quality. For example, intelligent \ac{ML} algorithms enhance image  resolution, correct noise and lighting artifacts, and automatically identify anatomical structures or abnormalities \cite{boucherit2025reinforced}. \ac{AI}-assisted image reconstruction can also restore diagnostic details lost due to environmental or hardware constraints, while predictive models optimize imaging parameters for each scenario. This intelligent feedback loop enables rapid, field-based decision-making without reliance on hospital infrastructure \cite{allegretta2020macro, iglesias2022quantitative,habchi2023ai}. \Acf{DL}, particularly \ac{CNN}-based architectures, plays a central role in \ac{PMI} image interpretation with performance approaching expert-level assessment \cite{dong2021feature,hwang2020smartphone}. Pre-trained models can continuously adapt to new imaging conditions through fine-tuning, improving robustness in variable field settings. Consequently, \ac{PMI} devices evolve from simple imaging tools into adaptive diagnostic platforms capable of autonomous decision support in emergencies \cite{habchi2025advanced, habchi2025machine}. \Ac{TL} enhances this adaptability by leveraging large-scale pretrained networks, reducing computational cost and training time. By reusing learned visual representations, \ac{TL} enables efficient domain adaptation to new organs, imaging modalities, or environmental conditions, significantly improving diagnostic consistency and generalization \cite{chen2022smartphone, zhu2020deep,vidal2021multi,habchi2024ultrasound, habchi2024deep}.  Building on the earlier discussion, \ac{PMI} requires \ac{MIQA} mechanisms that can (i) detect quality failures early, (ii) characterize diagnostically relevant distortions, and (iii) enable actionable responses (e.g., repeat acquisition, parameter adjustment, or escalation to expert review) prior to clinical interpretation. This motivates surveying \textit{AI-enabled} \ac{PMI} pipelines where learning improves not only interpretation but also acquisition via enhancement, artefact reduction, reconstruction support, and domain adaptation under portable constraints. The proposed review covers conventional \ac{ML} and \ac{DL} approaches, as well as advanced \ac{DL} paradigms—including \ac{TL}, Teansformers, and more —across the full workflow, and discusses key challenges such as domain shift, limited labeled data, and ensuring trustworthy \ac{AI} in safety-critical settings, among others.

\subsection{Existing review}

Existing reviews on \ac{PMI} can be broadly clustered by imaging modality and clinical domain; however, the majority adopt a technology-driven or application-centric perspective rather than a quality-centric one grounded in \ac{MIQA}. As a result, image quality is typically discussed implicitly—through diagnostic accuracy, usability, or hardware evolution—rather than explicitly formalized through distortion models, quality metrics, datasets, and validation protocols.

In \ac{PCT}, prior reviews primarily emphasize system evolution, portability, and bedside feasibility. For example, \cite{ce2024portable} surveys dynamic digital radiography for chest imaging, focusing on functional parameters such as motion and perfusion surrogates and highlighting clinical advantages in point-of-care settings. While clinically informative, this line of work does not formalize image degradation sources, objective or subjective quality assessment metrics, nor reproducible evaluation pipelines required for \ac{MIQA}. Similarly, \ac{PMRI} reviews predominantly focus on physics-driven trade-offs between \ac{SNR}, spatial resolution, contrast, and acquisition time, alongside mitigation strategies based on reconstruction, encoding, or hardware design. The low-field MRI perspective introduced in \cite{wald2020low} reframes image quality as “fitness-for-purpose,” which is particularly relevant for portable systems; nevertheless, it still lacks a unified treatment of \ac{MIQA} methodologies, benchmark datasets, distortion-aware evaluation, and standardized reporting practices under portable constraints.

In the context of \ac{PUS}, existing surveys concentrate largely on clinical feasibility, diagnostic accuracy, and deployment considerations in specific scenarios. The scoping review in maternal–fetal imaging by \cite{eggleston2022portable} maps device availability and gestational-age accuracy evidence, while \cite{shaddock2022potential} synthesizes barriers related to training, infrastructure, and rural or remote deployment. Comparative diagnostic performance against high-end \ac{US} systems is discussed in \cite{rykkje2019hand} and lung-specific applications are reviewed in \cite{haji2021hand}. Although image quality is occasionally addressed—such as in breast imaging evaluations \cite{ibraheem2022evaluation}—it is typically framed through qualitative assessments, operator dependence, or task-specific outcomes rather than through a comprehensive, modality-agnostic \ac{MIQA} framework incorporating objective metrics, subjective protocols, and reproducible benchmarking.

Across all modalities, a common limitation emerges: the absence of an integrated \ac{PMI}-oriented \ac{MIQA} review that systematically consolidates (i) objective and subjective quality assessment methodologies, (ii) distortion taxonomies inherent to portable imaging (e.g., noise, motion artifacts, low contrast, limited dynamic range, and electromagnetic interference), and (iii) publicly available datasets, evaluation protocols, and validation strategies tailored to real-world portable constraints. Furthermore, existing reviews rarely establish explicit links between image quality assessment and downstream clinical workflows, AI robustness, or deployment-oriented performance trade-offs.

Our proposed review addresses these gaps by providing a comprehensive and systematic analysis of image quality considerations in \ac{PMI}, with a specific focus on how acquisition constraints, environmental conditions, and device limitations affect diagnostic reliability. Rather than proposing a new \ac{MIQA} framework, this work critically surveys and organizes existing \ac{MIQA} approaches in the context of portable imaging, covering \ac{PMRI}, \ac{PUS}, and \ac{PCT} modalities under real-world operating conditions. The review explicitly examines both objective and subjective quality assessment strategies, highlighting their respective strengths, limitations, and suitability for portable environments. Specifically, the paper (i) analyzes the major sources of quality degradation in \ac{PMI}, including motion artifacts, noise, low contrast, geometric distortion, electromagnetic interference, and post-acquisition degradations introduced by processing or transmission; (ii) systematically reviews classical, \ac{ML}–based, and \ac{DL}–based \ac{MIQA} metrics, emphasizing their diagnostic relevance and limitations when applied to portable data; and (iii) consolidates publicly available datasets, evaluation protocols, and benchmarking practices used in current \ac{PMI}-related studies. In addition, the review surveys AI-driven enhancement, reconstruction, and quality prediction techniques—spanning ML, DL, transfer learning, and Transformers—that aim to compensate for degraded image quality in portable settings. The comparative positioning of existing surveys and the scope of the present review are summarized in Table~\ref{tab:CMPreviews}.

\begin{table}[!t]
\centering
\caption{Comparison of the proposed reviews with existing \ac{PMI}-based \ac{MIQA}-related surveys and reviews. A mark (\cmark) signifies that a specific area has been covered, whereas a mark (\xmark) denotes that it has not been addressed.}
\scriptsize
\begin{tabular}{ccccccccccccccccc}
\label{tab:CMPreviews} \\
\hline

\multirow{2}{*}{\textbf{Ref}} & \multirow{2}{*}{\textbf{Year}}& \multirow{2}{*}{\textbf{Paper}}  &\multirow{2}{*}{\textbf{QC}}&\multicolumn{3}{c}{\textbf{AI-PMI}}& \multirow{2}{*}{\textbf{Metrics}} & \multirow{2}{*}{\textbf{Dataset}} & \multirow{2}{*}{\textbf{PMRI}}& \multirow{2}{*}{\textbf{PUS}}& \multirow{2}{*}{\textbf{PCT}} & \multirow{2}{*}{\textbf{WCE}} & \multirow{2}{*}{\textbf{Apps}} & \multirow{2}{*}{\textbf{CH}}  & \multirow{2}{*}{\textbf{FDs}} & \multirow{2}{*}{\textbf{NRP}}\\

\cline{5-7}

& & & & \textbf{ML}& \textbf{DL}& \textbf{AD-DL} & & & & & & & & & &\\

\hline

\cite{ce2024portable} & 2024& Review  &\xmark & \xmark & \xmark & \xmark &\xmark & \xmark&\xmark&\xmark&\cmark&\xmark &\cmark&\cmark&\cmark&03\\

\cite{wald2020low} & 2020& Review &\xmark &  \xmark & \xmark & \xmark &\xmark & \xmark&\cmark&\xmark&\xmark&\xmark&\cmark &\cmark&\xmark&02\\

\cite{eggleston2022portable} & 2022&Review  &\xmark   & \xmark & \xmark & \xmark &\xmark &\cmark&\xmark&\cmark&\xmark&\xmark &\xmark&\xmark&\cmark&06\\

\cite{shaddock2022potential} & 2022&Review   &\xmark & \xmark & \xmark & \xmark &\xmark &\xmark&\xmark&\xmark&\cmark&\xmark &\xmark&\xmark&\xmark&16\\

\cite{rykkje2019hand} & 2019&Review   &\xmark & \xmark & \xmark & \xmark &\xmark &\xmark&\xmark&\cmark&\xmark&\xmark &\cmark&\xmark&\xmark&03\\

\cite{haji2021hand} & 2021&Review   &\xmark & \xmark & \xmark & \xmark &\xmark &\xmark&\xmark&\cmark&\xmark&\xmark&\cmark &\xmark&\xmark&05\\

\cite{ibraheem2022evaluation} & 2022&Review   &\xmark & \xmark & \xmark & \xmark &\xmark &\cmark&\cmark&\xmark&\xmark&\xmark&\cmark &\xmark&\xmark&06\\

\cite{wajid2020comparison} & 2020&Review   &\xmark & \xmark & \xmark & \xmark &\xmark &\xmark&\xmark&\cmark&\xmark&\xmark &\xmark&\xmark&\xmark&10\\

\cite{kravchenko2025low} & 2025&Review    & \xmark & \xmark & \xmark &\xmark &\xmark&\xmark&\cmark&\cmark&\xmark &\xmark&\cmark&\xmark&\xmark&11\\

\textbf{Our} & 2026 & Review & \cmark & \cmark & \cmark & \cmark & \cmark & \cmark & \cmark& \cmark& \cmark& \cmark& \cmark& \cmark& \cmark& 19\\
\hline
\end{tabular}
\begin{flushleft}
\textbf{Abbreviations:} \scriptsize{QS: Quality consideration; AD: Advanced; Apps: Applications; CH: Challenges;  FDs: Future directions;  NRP: Number of review paper.}
\end{flushleft}
\end{table}

\subsection{Contribution}

The purpose of this paper is to provide an overview of portable medical imaging technologies and methods. The focus is primarily on the various key tasks that make up \ac{PMI} systems, from the acquisition phase to the detection and classification stage, including pre-processing and post-processing. The aspect of image quality in the context of \ac{PMI} is also discussed in light of new advances in technologies and methodologies for better use of artificial intelligence-based solutions. The principal contributions of this work are summarized as follows:

\begin{itemize}
\item A detailed overview of \ac{PMI} devices, focusing on their types, principles, sensors, data processing, connectivity, and distortions impacting the quality of medical imaging.
\item A comprehensive overview of modern \ac{AI}-driven frameworks used in \ac{PMI}, including \ac{ML}, \ac{DL}, and advanced \ac{DL} models, with a detailed taxonomy and analysis of their core concepts for \ac{PMI}.
\item An exploration of \ac{MIQ} assessment, emphasizing its importance, subjective and objective evaluation methods, along with an overview of \ac{PMI} devices benchmark datasets and \ac{MIQ}  protocols for accurate analysis.
\item An overview of \ac{PMI} applications across various fields, including FPGA design, disease detection, image denoising, and segmentation, detailing their advantages, performance, strengths, and limitations.
\item A discussion on the research challenges and future opportunities in \ac{PMI}, focusing on technological hurdles, accuracy, efficiency, integration, and scalability, while exploring potential innovations to enhance medical imaging capabilities.

\end{itemize}

This review is designed for a diverse audience, including researchers, clinicians, and technologists in the fields of medical imaging, \ac{AI}, and healthcare. It specifically targets those focused on improving diagnostic accuracy, enhancing the efficiency of medical imaging processes, and advancing the integration of cutting-edge \ac{PMI} systems into clinical practice. This work is particularly valuable for individuals interested in leveraging \ac{AI}-driven technologies to optimize image quality, streamline workflows, and contribute to more precise and effective healthcare delivery. The subsequent sections are organized as follows: Section 2 outlines the review methodology employed for \ac{PMI} based on \ac{AI} and \ac{IQA} frameworks, Section 3 introduces the fundamentals of \ac{PMI}, Section 4 presents the taxonomy of \ac{AI}-based \ac{PMI}, Section 5 highlights the most \ac{MIQA} and relevant datasets, Section 6 covers \ac{PMI} applications, Section 7 discusses research challenges and future directions, and Section 8 provides a conclusion for this review.

\section{Review methodology}

\begin{figure}[pos=t]
\centering
\begin{minipage}[c]{0.54\textwidth}
    \centering
    \includegraphics[width=0.9\textwidth]{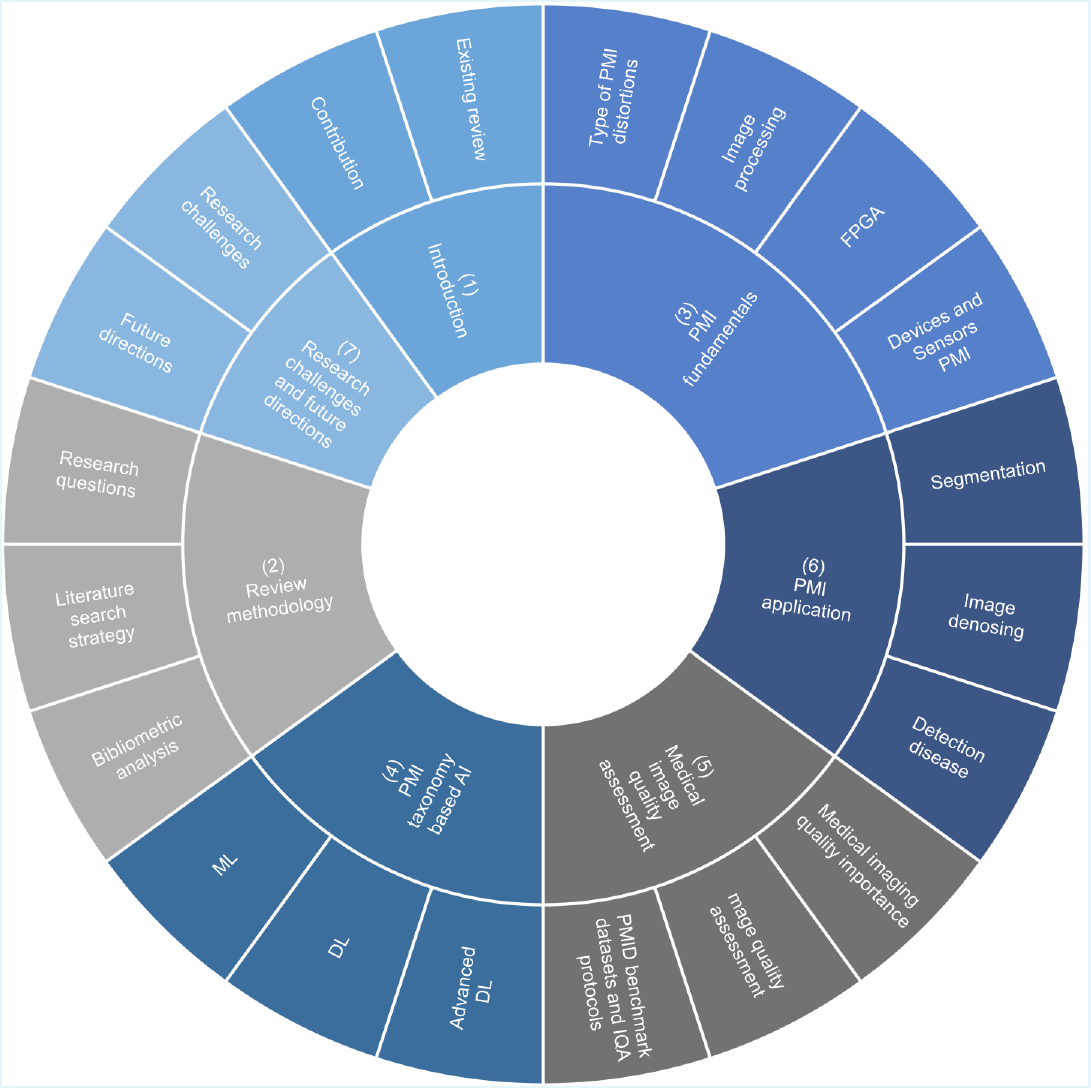}
\end{minipage}\hfill
\begin{minipage}[c]{0.36\textwidth}
    \centering
    \includegraphics[width=0.9\textwidth]{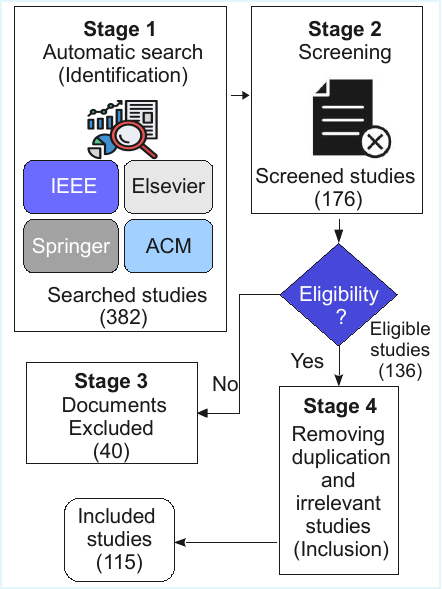}
\end{minipage}
\caption{An overview of how the review is structured and the criteria used for selecting studies. (a) Roadmap presenting the main sections covered in this review. (b) PRISMA flow diagram for the selection of articles.}
\label{figroadmapping}
\end{figure}

This section describes the methodology adopted to identify, screen, and analyze the literature included in this review. The aim is to ensure a transparent and rigorous survey process covering the fundamentals of \ac{PMI}, \ac{AI}-based frameworks, \ac{MIQA}, applications, and research challenges. Figure~\ref{figroadmapping}(a) presents the conceptual roadmap of the survey, including the main thematic sections addressed in this review.

To guide the analysis in a structured manner, five research questions are formulated in Table~\ref{tab:questions}. These questions define the main scope of the review and clarify the key objectives addressed throughout the manuscript. They cover the influence of portable operating conditions on diagnostic image quality, the limitations of conventional \ac{IQA} metrics, the role of \ac{AI} and \ac{DL} in real-time quality enhancement, the adaptability of advanced methods such as \ac{TL} and Transformers, and the development of clinically aligned and context-aware \ac{IQA} frameworks for \ac{PMI}.

\FloatBarrier
\begin{longtable}{p{3.5cm} p{6.2cm} p{7cm}}
\caption{Summary of Research Questions and Objectives.}
\label{tab:questions} \\

\hline
\textbf{Research focus} & \textbf{Research questions} & \textbf{Objectives / Expected outcomes} \\
\hline
\endfirsthead

\hline
\textbf{Research focus} & \textbf{Research questions} & \textbf{Objectives / Expected outcomes} \\
\hline
\endhead

\hline
\endfoot

\hline
\endlastfoot

\textbf{R1. Diagnostic quality in remote imaging} &
How do environmental and operational conditions (e.g., poor lighting, motion, humidity) affect diagnostic image integrity in \ac{PMI} systems? &
Identify the key environmental and device-related factors that degrade image quality and establish standardized criteria for diagnostic image integrity under field conditions. \\

\textbf{R2. Limitations of conventional \ac{IQA} metrics} &
Do existing \ac{IQA} metrics developed for controlled environments adequately assess \ac{PMI} image quality and diagnostic relevance? &
Critically evaluate classical \ac{IQA} metrics, such as PSNR and SSIM, on \ac{PMI} datasets and determine their sensitivity to distortions typical of portable imaging. \\

\textbf{R3. Integration of \ac{AI} and \ac{DL}} &
How can \ac{AI} and \ac{DL} models be adapted to assess and enhance image quality in real time for \ac{PMI}? &
Design and benchmark \ac{AI}-based \ac{IQA} frameworks integrating \ac{CNN}s, \ac{ViT}s, and hybrid models for real-time enhancement and quality prediction. \\

\textbf{R4. \ac{TL} and Transformers for \ac{PMI}} &
Can advanced \ac{DL} methods, such as \ac{TL} and Transformer-based techniques, improve the adaptability of \ac{IQA} models across diverse imaging modalities and environmental contexts? &
Develop adaptive \ac{IQA} models using advanced \ac{DL} methods to improve generalization across devices, clinical scenarios, and acquisition conditions. \\

\textbf{R5. Toward adaptive, clinically aligned \ac{IQA} frameworks} &
What design principles ensure that \ac{AI}-based \ac{IQA} frameworks align with clinical diagnostic needs and field usability? &
Propose a robust, context-aware \ac{IQA} framework tailored to \ac{PMI}, validated against expert evaluations to ensure diagnostic reliability and operational efficiency. \\

\end{longtable}

A systematic literature search was conducted across major scientific databases, including Scopus, Web of Science, ScienceDirect, SpringerLink, the ACM Digital Library, and IEEE Xplore. The search primarily focused on studies published during the last five years in order to capture recent and reliable advances in portable imaging and \ac{AI}-based quality analysis. Earlier works were also considered when necessary to provide foundational background on core concepts, benchmark datasets, and reference standards. In addition, a limited number of high-quality preprints, including works from arXiv, were consulted to reflect emerging developments.

The search process relied on keywords appearing in the title, abstract, and author keywords of candidate papers. The query was designed to retrieve studies specifically related to \ac{PMI}, medical imaging quality, and intelligent analysis methods, as follows:

\[
\begin{gathered}
\text{Selected papers} = \text{FROM (Abstract || Title || Keywords) SELECT} \\
\text{(References WHERE Keywords = (}\ac{PMI}\text{ || Medical imaging || Quality image) \& (}\ac{ML}\text{ || }\ac{DL}\text{))}
\end{gathered}
\]

where \texttt{\&} and \texttt{||} denote the logical operators AND and OR, respectively. To improve the relevance and rigor of the selection process, explicit inclusion and exclusion criteria were defined. The inclusion criteria retained studies that explicitly addressed the quality of medical images acquired using \ac{PMI} and that were published in indexed journals or conference proceedings. In contrast, studies were excluded if they did not focus on the targeted technological topics or if they referred to \ac{PMI} in unrelated or ambiguous contexts, such as non-medical engineering applications.

Figure~\ref{figroadmapping}(b) illustrates the PRISMA-based study selection process. From 382 initial records identified across the selected databases, 176 studies were screened, 40 were excluded, and 136 were assessed for eligibility, resulting in 115 studies included in the final review. These selected studies constitute the core body of literature analyzed in this survey.

\begin{figure}[pos=t]
\centering
\includegraphics[width=0.48\textwidth]{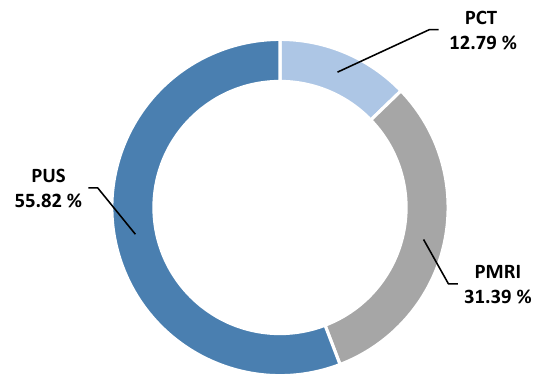}
\caption{Bibliographic statistics illustrating the distribution of research papers across PMRI, PCT, and PUS.}
\label{figstatistics}
\end{figure}

To further ensure the scientific quality of the review, the selected literature was examined using several indicators, including journal quartile ranking and database indexing. Highly relevant and influential studies were discussed in detail in the manuscript and comparative tables, whereas papers of more limited relevance were cited more selectively. Finally, Figure~\ref{figstatistics} presents the bibliometric distribution of the reviewed literature across the principal \ac{PMI} modalities, namely \ac{PMRI}, \ac{PCT}, and \ac{PUS}. Some references outside the 2020--2025 period were retained to support the introduction, compare prior reviews, and provide additional context on persistent challenges and future research directions.

\FloatBarrier

\section{PMI fundamentals}
This section introduces the fundamentals of \ac{PMI} by viewing portable imaging as an end-to-end \emph{sense--digitize--reconstruct} pipeline. As illustrated in Figure~\ref{figs_ct_mri_us}, a typical \ac{PMI} workflow starts from the clinical \emph{input} (patient scanning and protocol selection), proceeds through modality-specific \emph{energy/excitation} and \emph{signal generation}, then converges to a shared \emph{analog front-end} and \emph{A/D conversion} stage that turns physical measurements into digital samples. Finally, \emph{image reconstruction and post-processing} transform these samples into diagnostically useful images, which are displayed and often transmitted to hospital information systems.

Importantly, portability introduces tight constraints on power budget, thermal dissipation, shielding, and compute resources. These constraints shape sensor design (sensitivity, dynamic range, array geometry), acquisition strategies (sampling, synchronization, motion compensation), and reconstruction choices (speed vs.\ accuracy). The following subsections discuss the sensing hardware, the digitization and processing chain, connectivity requirements, and the main sources of accuracy loss and distortion that can affect \ac{PMI} data.

\subsection{Devices and sensors in \ac{PMI}}

\begin{figure}[pos=t]
\centering
\includegraphics[scale=0.68]{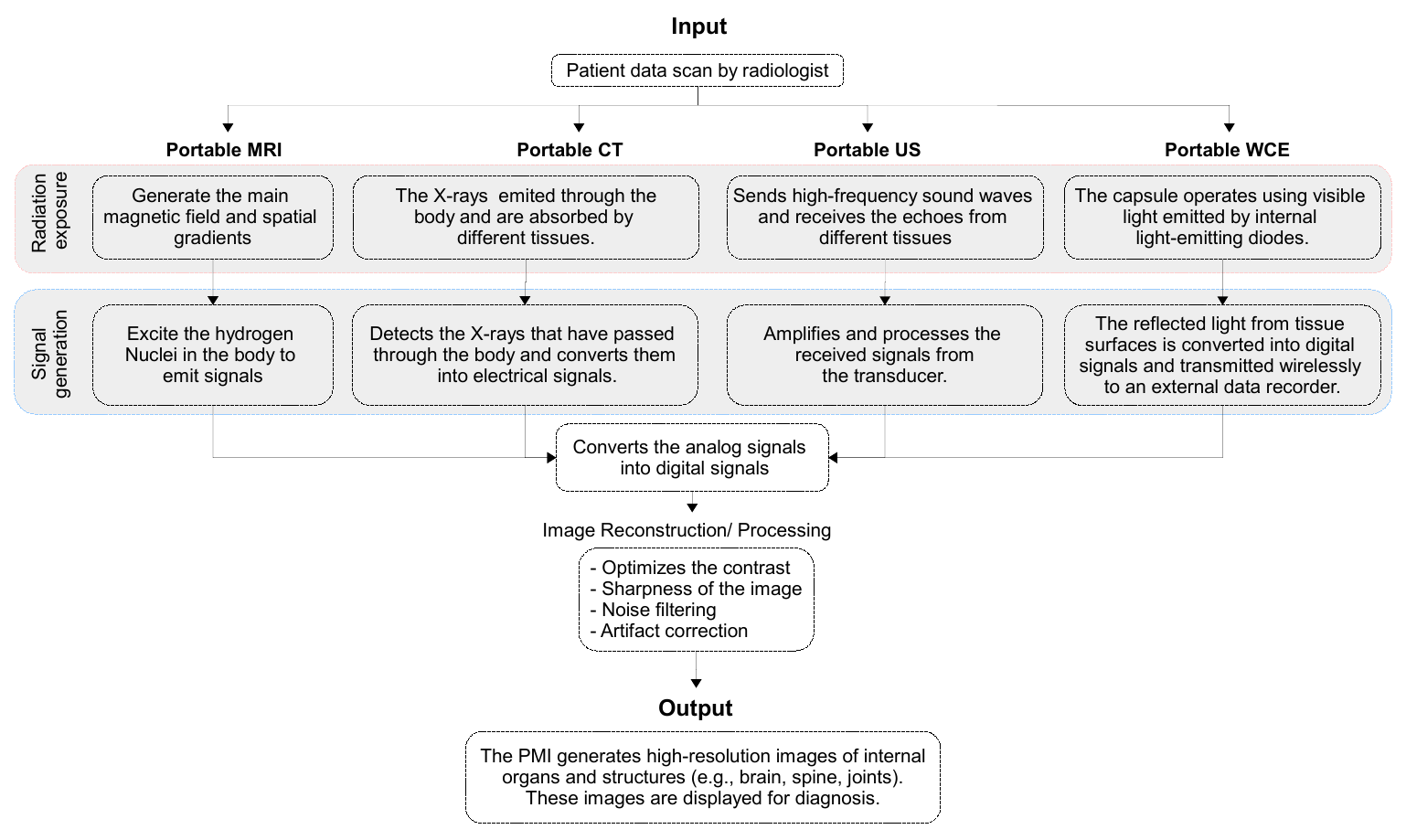}
\caption{Block diagram of  PMI working principle for PMRI, PCT, PUS, and WCE.} 
\label{figs_ct_mri_us}
\end{figure}

\ac{PMI} systems rely on advanced sensing technologies that enable rapid, on-site diagnostics in remote locations and emergency settings. Despite compact hardware constraints, image quality depends on optimized sensors and acquisition strategies. As shown in Figure~\ref{figs_ct_mri_us}, sensor performance should be understood within the complete pipeline: (i) energy delivery/excitation (x-rays, magnetic fields/RF pulses, or acoustic waves), (ii) signal generation and detection, (iii) conditioning and digitization (low-noise amplification, filtering, and \ac{ADC}), and (iv) reconstruction and post-processing (contrast optimization, noise filtering, and artifact correction). Accordingly, \ac{PMI} devices can be broadly classified into three categories:

\medskip
\noindent\textbf{(a) \Ac{PCT} devices:} In \ac{PCT} systems, low-power x-rays pass through the patient and are measured by detector sensors. A compact detector typically combines a scintillator (x-ray to light conversion) with photodiodes (light to electrical signals), followed by low-power readout electronics for amplification, filtering, and digitization. The resulting multi-view projections are reconstructed using methods such as filtered back projection or iterative algorithms to generate cross-sectional \ac{CT} images for point-of-care use \cite{jacobi2020portable, kandemir2019place}. As indicated in Figure~\ref{figs_ct_mri_us}, image quality and safety depend on exposure management, detector sensitivity and dynamic range, and artifact correction (e.g., scatter, beam hardening, and motion), with portable operation requiring careful synchronization and calibration. In this context, \cite{pande2016sensor} reports a hand-held optical coherence tomography system that integrates a motion sensor to track probe displacement and reduce motion-induced reconstruction errors.

\medskip
\noindent\textbf{(b) \Ac{PMRI} devices:} In \ac{PMRI} systems, portable operation requires generating the main magnetic field and gradients, exciting hydrogen nuclei with RF pulses, and detecting weak resonance signals for image formation (Figure~\ref{figs_ct_mri_us}). RF coils act as transmitters and receivers and largely determine the signal-to-noise ratio, while gradient coils encode spatial information. Since low-field \ac{PMRI} produces weak signals, low-noise preamplifiers, stable timing, and suitable \ac{ADC} stages are needed for reliable digitization before reconstruction from $k$-space measurements \cite{desroche2024feasibility, cooley2017design}. To improve robustness in non-ideal environments, \cite{yang2022active} combines RF sensing with electromagnetic interference (EMI) sensors for synchronous acquisition and noise suppression, while \cite{greer2019easily} reports a single-sided portable \ac{MRI} sensor integrating permanent magnets, planar RF coils, and printed gradient coils with shielding and interference monitoring for stable low-field imaging.

\medskip
\noindent\textbf{(c) \Ac{PUS} devices:} In \ac{PUS} systems, the \ac{US} transducer is the core sensor, using piezoelectric arrays to transmit acoustic waves and receive tissue echoes (Figure~\ref{figs_ct_mri_us}). Echoes are converted into electrical signals that are amplified and digitized, while compact arrays support beam steering and focusing. Portable processing typically includes time-gain compensation and filtering, followed by beamforming to produce real-time bedside images \cite{shaddock2022potential, qiu2017portable}. Beyond the transducer, several works integrate auxiliary sensors to improve acquisition: \cite{sobhani2016portable} uses an \ac{IMU} for probe orientation tracking during manual scanning, \cite{valente2021new} combines an \ac{IMU} with a pressure sensor for realistic \ac{US} simulation, \cite{paraska2024innovative} reports a handheld system supporting multiple modes (B-mode, Doppler, SE, ACM) with adaptive operation, and \cite{ning2024cable} highlights force and tracking sensors in robotic \ac{US} to control probe pose and contact for stable imaging.

\medskip
\noindent\textbf{(d) \Ac{WCE} devices:} \ac{WCE} systems consist of a self-contained ingestible capsule integrating a miniature complementary metal-oxide-semiconductor (CMOS) camera, light-emitting diodes (LEDs), a battery, and a radiofrequency transmission module (Figure~\ref{figs_ct_mri_us}). WCE does not involve ionizing radiation or externally applied energy fields; instead, internal LED illumination enables optical visualization of the gastrointestinal mucosa as the capsule passively advances through the digestive tract via peristalsis. Reflected light from tissue surfaces is captured by the camera sensor, converted into digital image signals, and transmitted wirelessly to external receiver sensors worn by the patient \cite{el2024accurate}. For example, In \cite{mitrakos2024pressurecap}, the sensor integrated into the capsule is a flexible pressure sensor arranged in a multi-point array around the surface of the device. It converts mechanical changes generated by the pressure exerted by the gastrointestinal wall into electrical signals that can be analyzed.

Across \ac{PCT}, \ac{PMRI}, and \ac{PUS}, the transition from physical sensing to digital imaging hinges on (i) robust analog conditioning (low-noise amplification, impedance matching, and anti-alias filtering), (ii) accurate digitization (sufficient \ac{ADC} resolution and sampling stability), and (iii) reconstruction/post-processing that improves contrast and sharpness while suppressing noise and correcting artifacts. Distortions can arise at each stage: exposure/excitation variability, sensor non-linearities and drift, motion during acquisition, electromagnetic/acoustic interference, and model mismatch in reconstruction. Consequently, portable systems increasingly rely on lightweight calibration routines, artifact-aware reconstruction, and quality-control indicators (e.g., signal-to-noise proxies or motion metrics) to maintain diagnostic reliability under constrained operating conditions.

\subsection{FPGA-based processing in PMIs}

\ac{FPGA} technology played a critical role in \ac{PMI} devices due to its ability to deliver high-performance processing with low power consumption. \ac{FPGA}s enabled real-time data processing and computationally intensive operations required in medical imaging modalities such as \ac{US}, \ac{CT}, and \ac{MRI}. Their reconfigurable architecture allowed customization for specific clinical applications, ensuring flexibility and scalability in portable systems. Furthermore, \ac{FPGA}s reduced reliance on bulky and power-intensive CPUs, making them well suited for mobile healthcare solutions where size, weight, and energy efficiency were essential. Consequently, they served as key enablers for portable, cost-effective, and energy-efficient imaging systems, particularly in resource-limited environments \cite{sanaullah2018real, boyle2024battery}. For example, \cite{ibrahim20171024} introduced a 1024-channel single \ac{FPGA}-based beamformer for portable 3D \ac{US} systems, addressing the computational constraints of conventional 3D \ac{US} platforms. Traditional systems suffered from high computational demands associated with delay calculations for volumetric reconstruction. The proposed architecture employed a single \ac{FPGA} consuming approximately 5W to manage 1024 channels, achieving high-quality volumetric imaging with enhanced processing speed. A novel delay calculation algorithm reduced computational complexity and improved scalability. Implemented on a Kintex UltraScale KU-040 \ac{FPGA}, the system achieved 53.2 volumes per second while satisfying portable power requirements, demonstrating feasibility for low-cost 3D \ac{US} applications. Similarly, \cite{xu2018fpga} presented a resource-efficient digital beamformer for portable \ac{US} imaging implemented on a single \ac{FPGA}. The 128-channel receiver employed dynamic focusing with coarse and fine delay processing, complemented by a polyphase filter to improve spatial and contrast resolution. The design supported 40 frames per second and incorporated dynamic power management techniques that reduced energy consumption. Efficient memory management enabled handling of extensive delay data, and processed outputs were wirelessly transmitted to an Android device for visualization. The system demonstrated improved resolution and reduced power usage compared to conventional solutions. Likewise, \cite{wang2024portable} developed a handheld \ac{US} system based on \ac{FPGA}-driven synthetic aperture imaging. By integrating multiple transmit and receive channels, the system achieved high-quality imaging with optimized power efficiency. The architecture incorporated automated delay calculation and segmented apodization for parallel beamforming, improving memory utilization and processing efficiency. Supporting B-mode, C-mode, and D-mode imaging, the prototype delivered competitive image quality relative to commercial handheld devices while maintaining lower power consumption, thereby enhancing portability and suitability for constrained clinical settings.

\subsection{Image processing and connectivity}

After acquisition, sensor data are processed by embedded units (e.g., microprocessors or FPGAs) to improve image quality and diagnostic utility. Common stages include noise reduction (e.g., spatial filtering or wavelet or bandelet-based denoising \cite{habchi2026adaptive}), contrast enhancement to better separate tissues, and dynamic-range enhancement to reveal details in dark or saturated regions. In \ac{PUS}, \cite{lo2021color} proposes an optimized color Doppler engine that combines an adaptive clutter filter with a blood--tissue discriminator and computational reductions (smaller processing area and randomized down-sampling), improving blood-to-clutter separation while reducing processing time. For mobile x-ray imaging, \cite{choi2021usefulness} shows that lowering mAs reduces dose while preserving readability when coupled with noise reduction. Finally, \cite{ruiz2025influence} reports that applying enhancement filters to handheld x-ray radiographs did not significantly change diagnostic accuracy for proximal caries detection, indicating that such filters can be used without compromising performance.

Wireless connectivity (e.g., Wi-Fi, Bluetooth, and 5G) enables rapid access, sharing, and remote transmission of medical images from \ac{PMI} devices, supporting telemedicine and cloud-based workflows \cite{chai2023experience, dai2021high}. By enabling secure image transfer and centralized storage/processing, clinicians can collaborate and consult across locations, which is particularly valuable in rural or underserved settings. However, wireless dependence may introduce latency and bandwidth variability, and it increases cybersecurity risks that can impact time-critical diagnostics and regulatory compliance.

In addition, \ac{PMI} devices integrated with the \ac{IoT} represent a cutting-edge fusion of healthcare technology and intelligent connectivity. Through \ac{IoT} architecture, these devices such as portable \ac{US}, x-ray, and \ac{MRI} scanners are equipped with embedded sensors, wireless modules, and cloud connectivity that enable real-time data collection, transmission, and analysis. \ac{IoT} networks link these imaging tools to hospital information systems, allowing seamless synchronization with electronic health records (EHRs) and facilitating remote monitoring and diagnostics by specialists from any location. The use of \ac{IoT} protocols, including Bluetooth low energy (BLE), and Wi-Fi, ensures low-latency communication for high-resolution image transfer and telemedicine applications. Advanced \ac{IoT} analytics further enable predictive maintenance, automated reporting, and \ac{AI}-assisted image interpretation. By connecting imaging devices within an \ac{IoT} ecosystem, healthcare providers can achieve improved efficiency, reduced diagnostic delays, and enhanced accessibility to imaging services, especially in remote and resource-limited environments \cite{krishna2016computer, rajendran2024portable}.

\subsection{Type of \ac{PMI} distortions}
Compared with conventional clinical systems, portable devices are more prone to in-capture degradations, including sensor noise, motion blur, defocus, low or non-uniform illumination, geometric lens artefacts, and modality-specific artefacts, which can directly affect diagnostic reliability. Post-capture degradations, such as compression artefacts, transmission errors, enhancement-induced side effects, and noise amplification, may further reduce quality during storage, wireless transfer, or on-device processing. Notably, \ac{PMI} quality is diagnosis-oriented rather than aesthetic, so even subtle distortions that obscure clinically relevant structures can lead to misinterpretation. Finally, the growing use of \ac{AI}-based analysis in portable imaging introduces additional vulnerabilities to adversarial or malicious perturbations, arising during acquisition (e.g., structured lighting or physical patterns) or after capture (e.g., pixel-level perturbations during processing or transmission) \cite{gillani2025distortion, beghdadi2020critical}.

\medskip

\noindent\textbf{Distortions in \ac{PMRI}:} \ac{PMRI} is prone to geometric distortion due to low magnetic field strength, compact magnet architectures, and increased $B_0$ inhomogeneity, which primarily deforms anatomy (often near boundaries) and can limit accurate visualization and quantitative use. In \cite{mateen2021low}, portable low-field brain \ac{MRI} exhibited reduced image quality relative to conventional scanners, mainly from lower \ac{SNR} and spatial resolution; the dominant artifact was geometric distortion driven by field inhomogeneity and compact magnet design, most visible at skull boundaries and peripheral brain regions, where contours were warped without producing signal voids, although clinically relevant findings (e.g., demyelinating lesions and cortical atrophy) remained identifiable. To address distortion and noise jointly, \cite{schote2024joint} proposes a physics-informed \ac{DL} reconstruction that estimates both the image and the $B_0$ field map within an unrolled optimization framework; field inhomogeneity is regularized via spherical harmonics to enforce smoothness, improving distortion correction and denoising (higher \ac{PSNR}, lower \ac{RMSE} across noise levels) while enabling rapid single-acquisition reconstruction for more reliable low-field portable \ac{MRI}. Complementarily, de Vos \emph{et al.} \cite{de2021design} mitigate distortion at the system level by redesigning a 50\,mT Halbach-based portable \ac{MRI} platform with improved low-frequency hardware, including a higher-linearity axial gradient coil, a four-channel phased-array receive coil, and a battery-powered three-axis gradient amplifier with enhanced filtering; the study also reports characterization of amplifier behavior, gradient-induced eddy currents, and a practical quality-control protocol to track stability, noise, and \ac{SNR}, supporting more consistent in vivo imaging in low-resource settings. Finally, \cite{ljungberg2025characterization} provides multicenter evidence on geometric fidelity by evaluating 64\,mT portable ultra-low-field scanners using a standardized phantom with fiducial markers across international sites; distortion, measured as fiducial displacement from known references in high-resolution in-plane directions, was small, spatially consistent, and highly reproducible, increasing modestly with distance from isocenter and showing minimal site-to-site and temporal variation, indicating reliable geometric accuracy for multicenter structural neuroimaging.

\medskip

\noindent\textbf{Distortions in \ac{PUS}:} \ac{PUS} devices are vulnerable to image distortion due to hardware miniaturization, limited channel counts, array sampling constraints, and susceptibility to electronic losses and interference. These factors can increase grating lobes, elevate sidelobes, induce aperture nonuniformity, and exacerbate motion-related artifacts, thereby degrading spatial resolution, contrast, and geometric fidelity in portable \ac{US} imaging. In \cite{fuller2008experimental}, a portable low-cost C-scan \ac{US} prototype shows that geometric distortion is strongly governed by array sampling and channel integrity: a $32\times32$ fully sampled 2-D array exhibited grating lobes because the element pitch exceeded the acoustic wavelength, producing pronounced off-axis artifacts and a degraded point spread function, while $6.74\%$ channel loss (Printed circuit board routing errors and intermittent connector contact) created nonuniform apertures that raised sidelobes, distorted the mainlobe, and yielded shift-variant gain visible as clutter, gaps, and reduced contrast; these effects dominated subtle beamforming differences.  Related portability constraints also arise in compact \ac{PMRI}, where undersampling, noise, and limited acquisition/transfer bandwidth can introduce severe artifacts and geometric inconsistencies; \cite{shin2017compressed} mitigates these effects using compressed sensing and advanced reconstruction, reporting that Bayesian learning--based reconstruction with cosine-domain sparsity most effectively suppresses distortion while preserving contrast and SNR, enabling accurate imaging even under aggressive subsampling. Finally, \cite{kording2018evaluation} demonstrates that integrating a portable Doppler \ac{US} gating device for fetal cardiac \ac{MRI} can preserve geometric fidelity: safety and compatibility tests showed no RF interference, no $B_1$ inhomogeneity, no geometric distortion, and no signal loss from the device/cabling, while phantom and in vivo studies produced stable gating and distortion-free cine images across field strengths, supporting reliable low-distortion fetal cardiac \ac{MRI}.

\medskip

\noindent\textbf{Distortions in \ac{PCT}:} Distortions in \ac{PCT} arise from factors like limited detector resolution, patient motion, beam hardening, and system miscalibration. These can cause artifacts, geometric distortions, reduced medical image clarity, and affect diagnostic accuracy in \ac{PCT}. For example in \cite{orchard2012plausibility}, the  distortion arises from geometric deformations, misalignments between acquisition and reconstruction geometries, and reduced x-ray cone angles. As the scanner bends, emitter and detector positions shift, increasing reconstruction errors with larger deformations. Geometric mismatches cause blurring and streaking, especially in high-contrast areas, while reduced cone angles or scanners placed too close lead to under-sampling and missing data, further affecting image quality. In addition, as claimed in \cite{wirth2004c}, the distortion in mobile C-arm-\ac{CT} is primarily caused by issues with image artifacts, positioning, and resolution. The study compares C-arm-\ac{CT} to conventional \ac{CT} and other 2D imaging methods for evaluating screw positioning in simulated treatments of talus neck fractures. The C-arm-\ac{CT} images showed significant artifact formation in many cases, leading to reduced diagnostic quality. The positioning of the foot during C-arm-\ac{CT} scanning was found to influence the distribution of these artifacts, with less artifact formation when the foot was positioned parallel to the orbital axis. Additionally, the resolution of C-arm-\ac{CT} was inferior to conventional \ac{CT}, particularly in diagnosing subtle misplacement of screws, but it performed better than 2D C-arm fluoroscopy. 

\medskip

\noindent\textbf{Distortions in \ac{WCE}:} Distortions in \ac{WCE} primarily result from motion-related effects, optical limitations, illumination variability, and wireless transmission. Because the capsule moves freely through the gastrointestinal tract, unpredictable rotation and peristalsis can cause motion blur and inconsistent viewing angles, while the wide-angle lens introduces peripheral optical distortion. Non-uniform LED illumination may generate glare, shadowing, or reduced contrast due to fluids and debris, and the miniaturized CMOS sensor limits spatial resolution and depth perception. Additionally, wireless signal attenuation and occasional frame loss can affect temporal continuity \cite{mehedi2023intelligent}.

\section{\Ac{AI}-based \ac{PMI} taxonomy}
This section surveys the diverse algorithms underpinning \ac{AI}-based \ac{PMI} image processing, covering \ac{ML}, \ac{DL}, and advanced \ac{DL}-based techniques ( \ac{TL}, and Transformer) used across portable modalities. It highlights the role of these methods in preserving and enhancing image quality, and discusses their effectiveness, robustness, and adaptability under the operational constraints and variability typical of real-world \ac{PMI} deployments. Table \ref{tab:AI_based_PMI} presents a summary of \ac{AI}-based \ac{PMI}, highlighting the used method, contributions, used data, best results and limitations.

\subsection{ML-based methods}
\ac{ML} in medical image analysis on mobile devices, such as smartphones, represents a significant advancement in improving healthcare in remote and resource-limited areas. This field relies on \ac{AI} algorithms to provide real-time, accurate analysis without the need for bulky equipment or continuous internet connectivity. By using lightweight and optimized models designed for mobile devices, these systems can offer precise diagnostics based on medical images, contributing to better healthcare access in communities lacking strong medical infrastructure. These systems are capable of operating locally on devices, reducing the need for constant internet connections and increasing response speed. Despite challenges related to privacy and training models on diverse data, this technology is a vital tool for expanding healthcare access and achieving accurate, timely diagnoses.

Several work have proposed and discussed the importance of applying \ac{ML} in this type of \ac{PMI}s. For example, in \cite{xu2024development}, the authors presents the development of a \ac{ML}-based classification method for assessing carotid plaques using portable 3D \ac{US} technology. The authors use several types of features extracted from 3D image volumes carotid volumetric stenosis rate, low-intensity rate, grayscale median, fractal dimension, and 3D gray level co-occurrence matrix properties from 3D \ac{US} images. Subsequently, these features are then fed into a \ac{SVM} classifier. The results indicate that the \ac{SVM} classifier, incorporating multiple image features, can effectively classify carotid plaque vulnerability and may serve as a promising tool for early diagnosis and screening using portable 3D \ac{US} devices. Thus, this method offers a viable solution for portable, efficient, and accurate plaque detection. Similarly,  \cite{hashir2023tinyml} presents a pioneering approach to brain stroke detection using a TinyML-based portable, low-cost microwave imaging system. The system integrates a compact setup combining a Vivaldi antenna, a low-cost vector network analyzer, and a single-board computer with TinyML to process the imaging data. This innovative system addresses the limitations of conventional stroke detection techniques, offering portability, cost-efficiency, and high accuracy. The features employed in this paper for the stroke detection system include scattering parameters and frequency domain to time series conversion. The dataset used for training the model consists of 300 head-imaging scans, which were pre-processed for simplicity. The model achieved a high testing performances. Therefore, this approach represents a significant advancement in affordable, real-time stroke diagnosis using compact technology.

\subsection{DL-based methods}
\begin{figure}[pos=t]
\centering
\includegraphics[scale=0.6]{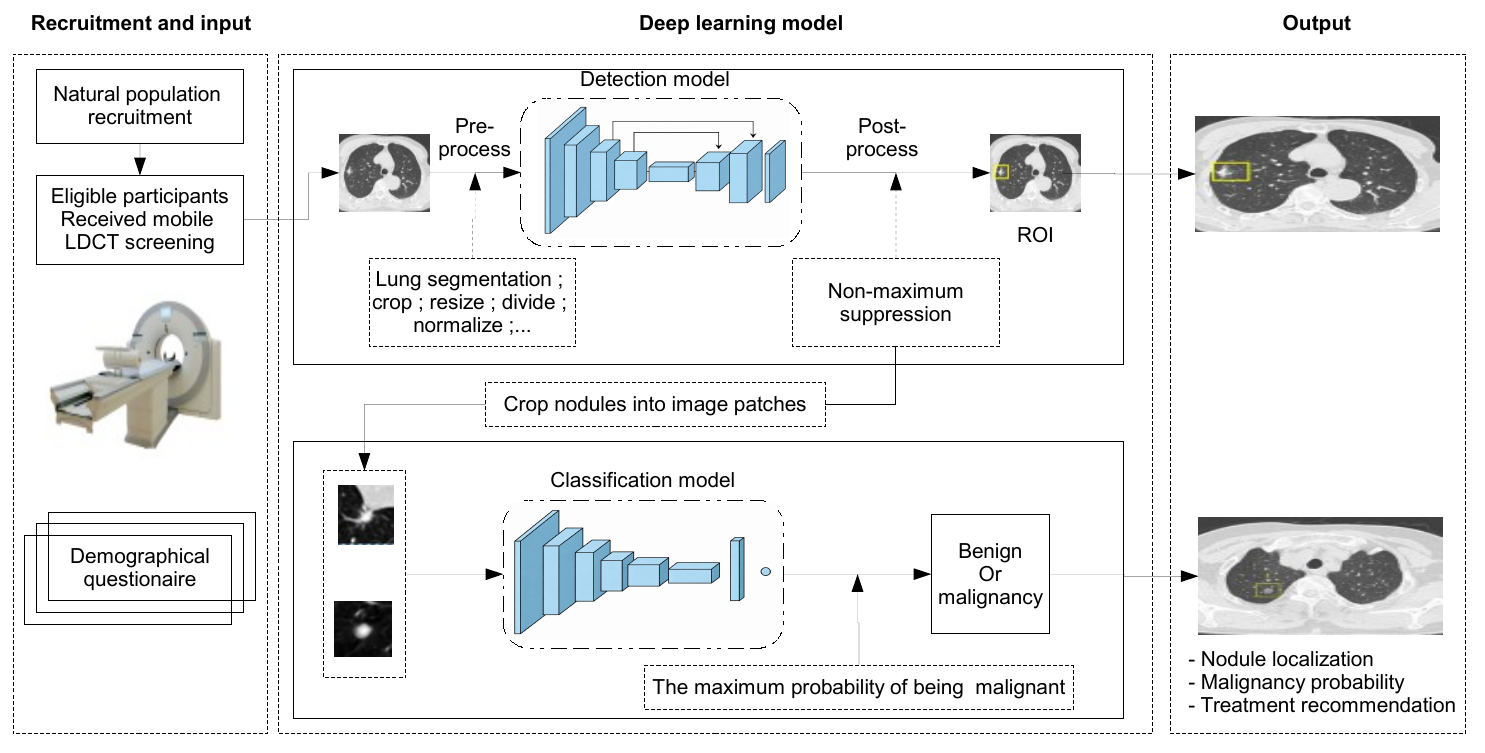}
\caption{Example of mobile low-dose \ac{CT} and \ac{DL} for lung cancer screening \cite{shao2022deep}. Figure illustrates the process of lung cancer screening using a mobile low-dose \ac{CT} (LDCT) device combined with a \ac{DL} model. Participants are first recruited and undergo mobile \ac{CT} scanning, along with completing demographic questionnaires. The \ac{CT} images are then processed by the \ac{DL} model, which operates in two stages: the detection model identifies potential nodules in the lung images, and the classification model analyzes these nodules to predict their malignancy. The output includes the localization of the nodules and their malignancy risk, aiding in early diagnosis and decision-making. } 
\label{ct1}
\end{figure}

\begin{figure}[pos=t]
\centering
\includegraphics[scale=0.65]{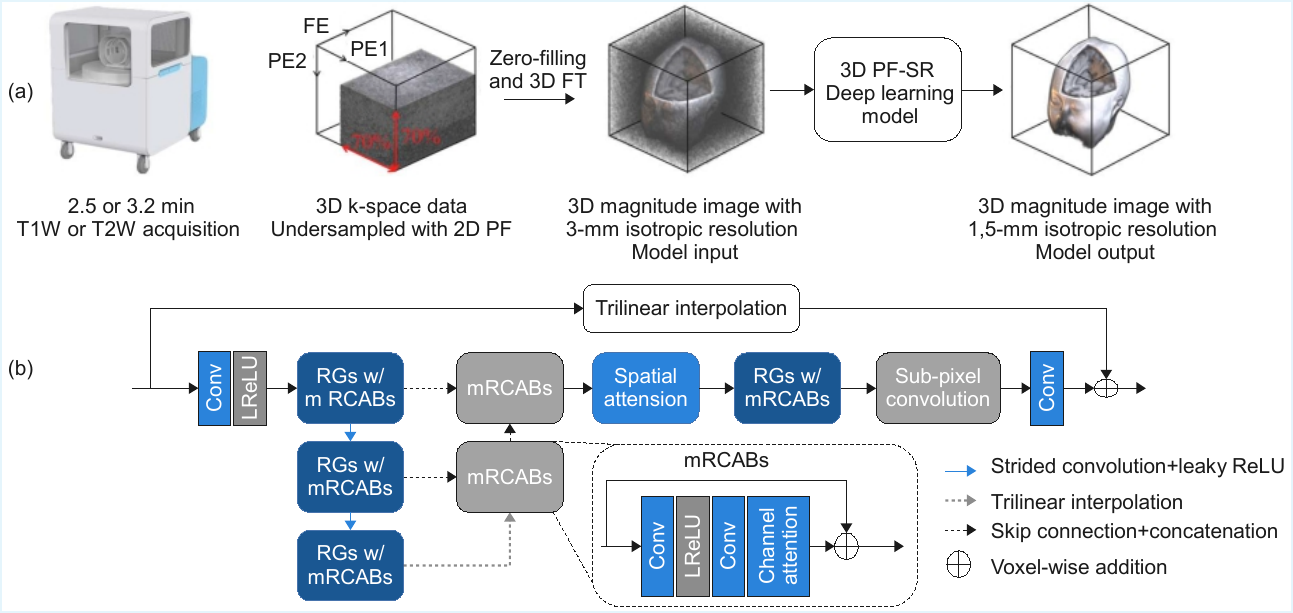}
\caption{Portable ultralow-field \ac{MRI} with \ac{DL} reconstructs 3D brain images \cite{man2023deep}. Illustrates the data acquisition and \ac{DL} reconstruction pipeline for accelerating ultra-low-field (ULF) isotropic 3D brain \ac{MRI} at 0.055 T. (a) Shows the acquisition process, where a single excitation (NEX) is used along with 2D partial Fourier (PF) sampling at a fraction of 0.7 in two phase-encoding directions. This reduces the scan time to 2.5 minutes for T1-weighted (T1W) and 3.2 minutes for T2-weighted (T2W) \ac{MRI}. After data acquisition, the raw 3D image data, with low \ac{SNR} and incomplete k-space, is reconstructed using the proposed \ac{DL} method. (b) depicts the overall architecture of the 3D PF-SR model, which employs residual groups (RGs), modified residual channel attention blocks (mRCABs), multiscale feature extraction, and spatial and channel attention to effectively restore fine brain structural features, suppress noise, and enhance spatial resolution to 1.5-mm isotropic.}
\label{mri1}
\end{figure}

\begin{figure}[pos=t]
\centering
\includegraphics[scale=0.6]{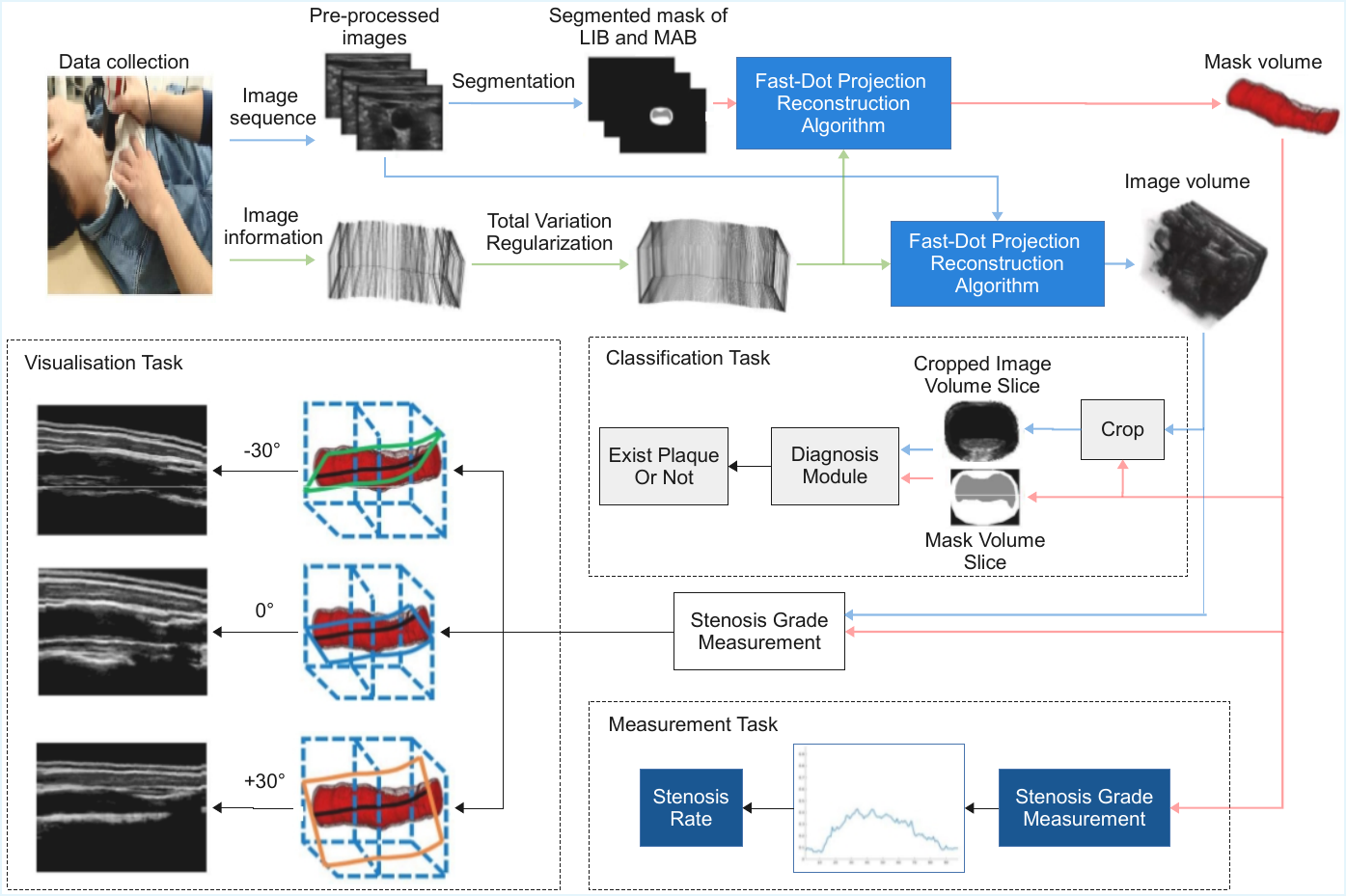}
\caption{Example of 3D \ac{US} imaging pipeline for carotid atherosclerosis using U-Net segmentation and regularized Fast-Dot Projection algorithm \cite{li2023automatic}. Figure illustrates the pipeline of the proposed technique for diagnosing carotid atherosclerosis using a portable freehand 3D \ac{US} imaging system. The top row showcases the data acquisition process, which involves collecting 2D transverse images along with positional information from the 3D \ac{US} device. These images are segmented using a U-Net algorithm, and 3D volume reconstruction is carried out using the Fast-Dot Projection (FDP) method with position regularization. The bottom row details the three main diagnostic tasks: (1) Visualization Task: Creation of longitudinal images at different angles to enhance visualization of the carotid artery, (2) Classification Task: Determining the presence of plaques in the carotid artery, and (3) Measurement Task: Quantifying the stenosis rate of the artery. The approach enables automatic diagnosis and measurement, making the system more accessible and reducing dependence on clinician experience.}
\label{us1}
\end{figure}

\ac{DL} in medical imaging uses multilayer neural networks, especially \ac{CNN}s, to analyze images in real time and support medical diagnosis. To run these models on low-resource devices, techniques such as pruning, quantization, and weight sharing are used to reduce energy and memory use while maintaining accuracy. Image quality is very important, as clear images with good contrast and low noise improve diagnostic performance. Therefore, simple image enhancement and preprocessing methods are applied to ensure reliable and real-time analysis, even on portable devices and in resource-limited environments \cite{kimberly2023brain,munew}. The three examples of \ac{PMRI}, \ac{PCT}, and \ac{PUS} based on \ac{DL} are shown in Figure \ref{ct1}, \ref{mri1}, and \ref{us1}.

\subsubsection{CNN models}

\Ac{CNN}-based approaches constituted the earliest and most widely adopted \ac{DL} paradigm for \ac{PMI}, owing to their strong capability to learn hierarchical spatial features directly from noisy and low-quality data. Within this category, existing studies primarily focused on image enhancement, geometric normalization, diagnostic classification, and computational efficiency under resource constraints.

\textbf{Noise suppression:}  
\ac{PUS} systems were particularly affected by electronic noise, limited aperture sizes, and hardware constraints, which severely degraded image quality and compromised diagnostic reliability. As a result, denoising approaches that explicitly preserved diagnostically relevant features became a central research focus in \ac{PUS} imaging. In light of this, the paper \cite{ma2021edge} presents the edge-guided denoising \ac{CNN} (EDCNN), designed to improve \ac{US} images from portable handheld devices. The model integrates canny edge detection to preserve edge information during denoising. EDCNN modifies existing denoising \ac{CNN} (DnCNN) and image restoration \ac{CNN} (IRCNN) architectures, incorporating edge-preserving techniques into the training process. Similarly, \cite{dong2021feature} proposed the \ac{FDCNN}, which used guided back-propagation, a technique from explainable \ac{AI} for explicit feature extraction combined with a hierarchical noise modeling strategy, where noise is selectively added to non-feature areas, ensuring that critical features are not corrupted. Additionally, the denoised images are fused with preserved features using Laplacian pyramid fusion, which ensures smooth integration of the denoised regions and retains important image details. In contrast, \cite{liftycnn} adopted a hybrid preprocessing and \ac{DL} strategy aimed at improving handheld \ac{US} image quality. The method combined median filtering, histogram equalization, and unsharp masking to enhance contrast, suppress noise, and sharpen anatomical details before refinement using \ac{CNN}s. Rather than explicitly modeling feature importance, the network learned to improve texture fidelity through supervised learning.

Despite their effectiveness, feature-preserving denoising methods often introduced additional computational overhead due to explainability mechanisms and multi-scale fusion architectures. Moreover, most existing studies relied on limited or device-specific datasets, raising concerns about generalization across probes, imaging protocols, and anatomical regions in real-world \ac{PUS} deployments.

\textbf{Normalization and scan conversion:}  
Beyond noise-related degradation, mobile \ac{US} imaging was also affected by geometric distortions and field-of-view (FOV) dependency, which hindered the performance and robustness of \ac{DL} models trained on heterogeneous acquisition geometries. The study in \cite{lee2021reverse} addressed this challenge by proposing a reverse scan conversion framework that enabled \ac{CNN}s to operate directly on pre–digital scan-converted ultrasound data. By removing FOV dependency, the approach improved classification accuracy across different probe configurations and imaging geometries. Additionally, the authors introduced a frame-asynchronous processing pipeline that allowed real-time image reconstruction and deep network inference to proceed concurrently, making the method suitable for latency-sensitive mobile ultrasound applications.

While reverse scan conversion effectively mitigated geometric bias, it introduced additional preprocessing stages that increased system complexity. Furthermore, the approach assumed access to raw or pre–scan-converted data, which was not always available in commercial handheld ultrasound systems.

\textbf{Imaging-based diagnosis:}  
In parallel with ultrasound imaging, portable radiography systems gained increasing attention, particularly for rapid point-of-care diagnosis in emergency and pandemic scenarios where conventional imaging infrastructure was unavailable. The work of \cite{de2020deep} presented a comprehensive \ac{CNN}-based framework for analyzing chest x-ray images acquired from portable devices, with a specific focus on COVID-19 detection. The proposed system employed three complementary deep models to handle binary and multiclass classification tasks, distinguishing between normal, pathological, and COVID-19 cases. Despite the reduced image quality inherent to portable radiography, the framework achieved robust diagnostic performance, highlighting the potential of portable x-ray systems to support clinical decision-making and patient triage in real-world settings.

Although promising, most portable x-ray studies relied on curated datasets that did not fully capture real-world variability in acquisition conditions, patient positioning, and disease prevalence. Generalization to unseen clinical environments and rare pathologies therefore remained an open challenge.

\textbf{On-device and medical image classification:}  
To address the strict energy, memory, and latency constraints of portable medical imaging platforms, a distinct line of research focused on computationally efficient learning paradigms that enabled real-time, on-device inference without specialized accelerators. The study by \cite{hou2020device} proposed a Saak Transform--based subspace learning framework combined with a feed-forward \ac{CNN} design that eliminated the need for back-propagation. Feature extraction was followed by ensemble classifiers, primarily Random Forests, with additional evaluations using SVM, KNN, and Na\"ive Bayes. Implemented on a Raspberry Pi~3B embedded platform, the system achieved competitive performance on large-scale chest x-ray datasets such as Open-I and NIH ChestXray14. This work demonstrated the feasibility of accurate medical image classification on low-power edge devices, offering an attractive alternative to conventional deep learning pipelines for portable and point-of-care imaging.

While highly efficient, subspace learning and shallow ensemble methods often lacked the representational capacity of modern deep architectures, particularly for complex or subtle pathological patterns. Balancing computational efficiency with diagnostic accuracy therefore remained a key challenge for future on-device medical imaging systems.

\subsubsection{GAN-based methods}
\Ac{GAN}-based methods in \ac{PMI} enhance image quality by generating high-resolution images from noisy or low-quality scans. They are used for denoising, super-resolution, and image reconstruction, improving diagnostic accuracy. These methods help preserve critical features, enabling better segmentation, feature extraction, and domain adaptation for more accurate decision-making in resource-limited settings. For example, the article \cite{jimenez2024branet} introduces BraNet, a mobile application designed for the classification of breast images, leveraging \ac{DL} algorithms for enhanced breast cancer diagnosis. The application, developed using React Native for iPhone operating system and android, integrates advanced \ac{ML} models, including SNGAN for synthetic image generation, segment anything model for segmentation, and ResNet18 for classification. Moreover, its ability to process both mammography and \ac{US} images. In addition, In \cite{zhou2019image}, the authors address the significant challenge of poor image quality in hand-held \ac{US} devices. They propose a novel two-stage \ac{GAN} model combining a U-Net network for structure extraction with a \ac{GAN} for fine detail and speckle enhancement. Although this model is useful in improving the quality of \ac{PMI} and preserving important diagnostic features, it still suffers from issues such as high computational resource consumption during training, dependence on large and high-quality datasets which may be difficult to obtain, along with the risks of overfitting and mode collapse, which could affect the diversity and quality of generated images.

\subsection{Advanced \ac{DL} algorithms applied to  PMI}
This section surveys advanced \ac{DL} frameworks in \ac{PMI}, including \ac{TL}-based models, and Transformer architectures, highlighting their respective roles in tackling data scarcity problem,  and long-range contextual modeling.

\subsubsection{TL-based methods}
\ac{TL} for \ac{PMI} devices leveraged models pre-trained on large-scale generic datasets and adapted them to task-specific medical applications. This strategy reduced training requirements and computational burden. The final network layers were customized for clinical tasks such as disease detection and optimized for deployment on resource-constrained platforms through pruning, quantization, and compression. As a result, real-time and accurate diagnostics became feasible on memory- and energy-limited devices, including smartphones and wearables, without cloud dependence. Thus, \ac{TL} facilitated advanced diagnostic capabilities in data-scarce and computationally constrained environments, enhancing the practicality of \ac{PMI} systems. Vidal et al. \cite{vidal2021multi} proposed a two-stage \ac{DL} framework for lung segmentation in chest X-rays acquired during the COVID-19 pandemic. Due to reduced image quality and limited training samples, knowledge was first transferred from a model pre-trained on brain \ac{MRI} images to general chest radiographs, and subsequently adapted using \ac{CT}-based data. This approach enabled robust segmentation under constrained imaging conditions. Kikkisetti et al. \cite{kikkisetti2020deep} applied a pre-trained ImageNet \ac{CNN} to classify COVID-19 and other pneumonia types from chest radiographs. Likewise, \cite{zhu2020deep} utilized a VGG16-based architecture to stage COVID-19 severity, reporting strong agreement between automated predictions and expert radiologist assessments. Similarly, \cite{chen2022smartphone} developed a smartphone-based diagnostic system for middle ear diseases using a lightweight \ac{CNN} adapted from a large otoendoscopic dataset. The model achieved high accuracy across ten conditions and demonstrated performance comparable to otolaryngology specialists, while enabling real-time, privacy-preserving diagnosis on mobile devices.

Despite its advantages, \ac{TL} presented limitations. Performance degradation could occur when source and target data distributions differed significantly. Moreover, adaptation on small datasets sometimes affected generalization, particularly when learned representations were not fully transferable to the new task.

\subsubsection{Transformer-based methods}

\begin{figure}[pos=t]
\centering
\includegraphics[scale=0.8]{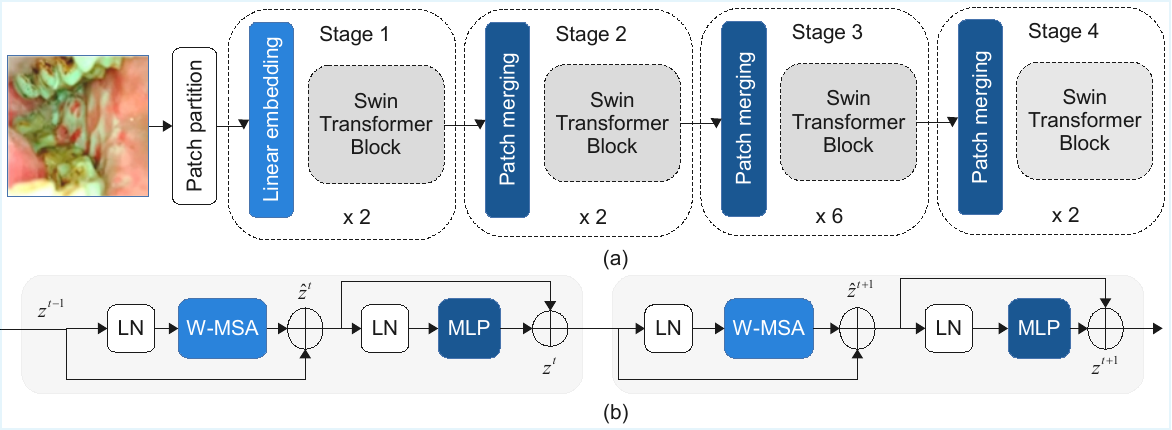}
\caption{Example of classification of mobile-based oral cancer images using the
\ac{ViT} and the swin transformer \cite{song2024classification}. (a) Shows the processing image patches through Swin Transformer blocks, reducing token numbers progressively. (b) Illustrates the Swin Transformer block, highlighting the multi-head self-attention mechanism, which processes patch tokens through transformations and passes them through a Multi-layer Perceptron head for classification. The use of a shifted window attention mechanism improves efficiency, making it highly effective for tasks like oral cancer image classification.}
\label{transformer}
\end{figure}

The \ac{ViT} is a groundbreaking model for \ac{PMI}, excelling in tasks like classification, segmentation, and anomaly detection. Unlike traditional \ac{CNN}s, \ac{ViT} uses self-attention mechanisms to capture global context, modeling long-range dependencies across the image. This enables more accurate image segmentation, crucial for tasks such as organ or tumor identification in medical imaging (See Figure \ref{transformer}). Optimized versions of \ac{ViT} offer real-time, accurate diagnostics on resource-constrained devices, ensuring high performance in environments with limited computing power. For example, the paper \cite{karageorgos2023transformer} explores the use of a \ac{ViT}-based architecture in the classification of benign and malignant breast lesions in \ac{US} images, particularly those obtained from handheld devices.  The study introduces a novel metric, the classification inconsistency rate, to quantify model uncertainty, which allows for more reliable predictions by excluding uncertain results. Although the Transformer model in \ac{PMI} is beneficial for capturing long-range dependencies and analyzing global context, it still suffers from high computational resource consumption, requiring significant processing power for training, which can be a challenge for portable devices with limited capabilities. Additionally, the need to optimize attention mechanisms in lightweight versions of Transformers may impact performance accuracy in resource-constrained environments. In addition, The study \cite{song2024classification} evaluates the effectiveness of \ac{ViT} and Swin Transformer models for mobile-based oral cancer image classification. By comparing these models with traditional \ac{CNN}s like VGG19 and ResNet50, the Swin Transformer outperforms the others, demonstrating its potential to improve oral cancer detection. The study highlights how transformer-based architectures can enhance image analysis, particularly in resource-limited settings. Data augmentation techniques were employed, and \ac{TL} improved performance, indicating that these models have significant potential for large-scale and effective oral cancer screening.

\begin{scriptsize}
\begin{longtable}{m{0.3cm} c m{1.5cm} m{4.5cm} m{1.2cm} m{1.2cm} m{1.5cm} m{3cm} c}
\caption{Summary of state-of-the-art \ac{AI}-based \ac{PMI} techniques, enabling easy visual comparison among the different proposed methods.}
\label{tab:AI_based_PMI} \\

\toprule
Ref & Year & Used method & Contributions & Used Data & Image modality & Best result & Limitations & Device \\
\midrule
\endfirsthead

\multicolumn{9}{c}%
{{\bfseries \tablename\ \thetable{} -- Continued from previous page}} \\
\toprule
Ref & Year & Used method & Contributions & Used Data & Image modality & Best result & Limitations & Device \\
\midrule
\endhead

\midrule 
\multicolumn{9}{r}{{Continued on next page}} \\
\bottomrule
\endfoot

\bottomrule
\endlastfoot

\cite{xu2024development} & 2024 & {SVM} & Developed a {ML}-based classification method using portable 3D {US} for carotid plaque detection. &- & 3D {US} & Acc:73.00\% \newline Sen: 75.00\% \newline Spe: 74.00\%.&Limited by feature extraction process. &{PUS}\\ \hline

\cite{wang2024deep} & 2024 & DeepLabv3, ResNeSt50 and DenseNet121  & Developed a {CAD} system using {DL} to localize nasogastric tubes and detect malposition on portable chest x-rays. & NTUH-20 & X-rays  & AUC:99.80\% \newline Sen: 98.80\% \newline Spe: 98.90\%& Imbalanced dataset and the testing datasets only included images obtained in emergency departments.&{PCT}\\
\hline
\cite{zhu2020deep} & 2020 &{CNN} and {TL} &Used {DL} and {TL} to accurately stage the severity of COVID-19 lung disease based on portable chest radiographs. & COVID-19 Chest x-rays&X-rays & MAE: 08.50\% &Small sample size and only one dataset. &{PCT}\\
\hline
\cite{hashir2023tinyml} & 2023 & TinyML & Introduced a TinyML-based portable, low-cost microwave imaging system for brain stroke detection. &Head-Microwave & Microwave  &  Acc:98.92\%  & Requires large datasets for more accurate training.&MWHI\\
\hline
\cite{hou2020device} & 2020 & Saak Transform-based {CNN}&A subspace weakly supervised model for chest x-ray classification, optimized to integrate {AI} on mobile devices and low-power platforms like Raspberry Pi.&Chest-Xary14 & x-rays images &Acc: 90.71\% &Requires ensemble methods for improved accuracy. &{PCT}\\
\hline
\cite{jimenez2024branet} & 2024& SNGAN \newline SAM \newline ResNet18 & Developed BraNet, a mobile app for 2D breast image classification using {DL}. & BUSI & {US}  & Acc: 94.70\% & Limited to two classes. &{PUS}\\
\hline
\cite{zhou2019image} & 2019 & U-Net + Residual {GAN} & Proposed a two-stage {GAN} combining U-Net for structure extraction and {GAN} for speckle/detail preservation, enhanced by gradual {TL} from plane-wave {US} data. & Simulation data& US images& PSNR: 23.17dB \newline SSIM: 0.55\newline BRISQUE:28.60 &Long training times. &{PUS}\\
\hline
\cite{kikkisetti2020deep} & 2020 & {CNN}-{TL} & Developed a {AI} system to classify COVID-19, bacterial pneumonia, non-COVID viral pneumonia, and normal portable chest x-rays. &COVID-19 pCXR & X-rays & Acc: 88.00\% \newline Sen:91.00\% \newline Spe: 93.00\%&Small sample size and potential selection bias. &{PCT}\\
\hline
\cite{bian2024quantitative} & 2024 & SCUNet & Developed a {DL}-based super-resolution model to improve ischemic lesion detection in portable low-field {MRI}. & SynthMRI & {MRI} & Sen: 89.00\% \newline Spe:91.30\% \newline P<0.001 & Potential dependence on quality of low-field-{MRI}. &PMD\\

\hline
\cite{deng2024portable} & 2024 & Multimodal {DRL}  & Developed a portable robotic {US} system with multimodal {DRL} for autonomous scanning. & -& {US} & SSIM: 76.40\% \newline NCC: 97.70\% & Dependency on target images for training. &{PUS}\\
\hline
\cite{zemi2024assessing} & 2024 & SSDLite320+ \newline MobileNetV3 & Explores the feasibility of integrating {AI} algorithms into {POCUS} for breast cancer detection & Phantom data& Breast phantom & 14.71 FPS \newline Time: 0.068 s &The frame rate of some models did not meet the 15 FPS requirement. &PMW\\
\hline

\cite{cobenas2023diagnostic} & 2023 & {AI}-B Alg. & Evaluated {AI}-A, B and C algorithms for detecting SARS-CoV-2 pneumonia on {PCT}. & COVID19& X-rays  & AUC:73.00\%& One hospital involved. &{PCT}\\
\hline
\cite{davidov2023incorporating} & 2023 & {ML} & Developed a {ML} model combining thermal imaging and image processing to detect metabolic dysfunction-associated steatotic liver diseas. & -& Thermal& Acc:72.00\%&Require more diverse data for better generalization. &PIT\\
\hline
\cite{karageorgos2023transformer} & 2023 & {ViT}& The study employs a vision transformer model for discriminating benign and malignant lesions in handheld {US} &- & {US} & Acc: 94.4\%, \newline AUC:96.00\%&. Limited to handheld {US} images.&{PUS}\\
\hline
\cite{hossain2022deep} & 2022 & YOLOv5l & Introduces a {DL}-based YOLOv5l model integrated into a portable MWHI for automatic brain tumor detection and classification. & MWHI& RMW  & Acc: 96.32\% \newline Pre: 95.17\% \newline F1: 95.53\%.& Only phantom images were used and real clinical data not tested.&RMW\\\hline

\cite{sheth2020first} & 2020 & U-Net  & A low-field {MRI} solution for acute brain injury using {AI} to assess mass effect. {AI}-based analysis showed strong correlation with human annotations of midline shift. & -& Brain & p: 0.01&Limitations in correlation with National Institutes of Health Stroke Scale. &{PMRI}\\
\hline
\cite{perez2020lightweight} & 2020 &MFQ-Net &Assessing eye fundus image quality with significantly fewer parameters than state-of-the-art models, optimized for mobile devices. & Kaggle DR & Diabetic Retinopathy images& Acc:91.10\% \newline  Sen:94.21\% \newline Spe:88.53\% & Limited to fundus image quality assessment. &SPH\\
\hline

\cite{dong2021feature} & 2021 & {FDCNN} & Enhancing portable {US} image quality by effectively denoising while preserving critical diagnostic features using explainable {AI}. & Private data& Clarius  & SSIM:0.9801 \newline PSNR:40.31dB & The model requires high computational power for training.&{PUS}\\
\hline
\cite{hwang2020smartphone} & 2020 & MobileNet& {AI}-based decision-making for detecting age-related macular degeneration in OCT images using {DL}. & Private data & OCT  & Acc: 90.02\%; \newline Sen: 92.51\%, \newline Spe: 85.93\%& Small dataset. &SPH\\
\hline
\cite{chen2022smartphone} & 2022 & {CNN}+{TL}& Developed a smartphone-based {AI} system using {TL} for diagnosing middle ear diseases. & Private data& Otoendo- scopic &Acc:97.60\% & Model performance may vary with low-resolution images. &SPH\\
\hline
\cite{vidal2021multi} & 2021 & U-Net {CNN}+{TL}&Adapt a pre-trained {MRI} model for lung segmentation in low-quality {PCT} images from COVID-19 patients. &Private data & X-ray &Acc:98.01\% \newline Sen:94.70\% \newline Spe:99.06\% & Artifacts due to low-quality images.&{PCT}\\
\hline
\cite{lee2021reverse} & 2021 & CNN &A reverse scan conversion method to enhance {CNN} accuracy for {US} image classification, using {DL} on mobile devices for real-time imaging. & Private data & {US} &Acc:88.79\% \newline Sen:74.86\% \newline Spe: 87.10\% & Different training and inference image processing.&{PUS}\\
\hline
\cite{de2020deep} &2020 & DenseNet-161& An automatic {DL} methodology for COVID-19 detection in chest x-ray images from portable devices. & Private data& Chest x-ray  & Acc:90.27\% & Some cases of mis-classification, the images present lower qualit and the complex characteristics
that may be present in image.&{PCT}\\
\hline
\cite{yang2022automatic} & 2022 & RefineDet & DL-based diagnostic model for knee osteoarthritis severity assessment on unpreprocessed radiographs. & Private data& X-rays  & Acc: 97.70\% \newline AUC:90.00\% & Potential errors in unpreprocessed images and limitations in detecting poor images.&{PCT}\\
\hline
\cite{ieracitano2022fuzzy} & 2022 & CovNNet & A framework to differentiate between Covid-19 and non-Covid-19 interstitial pneumonias using portable chest x-rays. & Private data& Chest x-ray& Acc:81.00\%, \newline  Sen:82.50\% \newline Spe: 78.60\% &Small dataset size and variable image quality. &{PCT}\\
\bottomrule
\end{longtable}
\end{scriptsize}

\ac{ML} and \ac{DL} methods have achieved significant success in portable imaging; however, their reliance on large annotated datasets and computationally expensive training limits real-world deployment. In contrast, \ac{TL} shows better adaptability but suffer from interpretability issues and domain bias. Transformer-based architectures, though promising, remain underexplored in \ac{PMI} and demand optimization for resource-constrained environments.

\section{Medical image quality assessment and datasets}

\subsection{Medical imaging quality importance}

\begin{figure}[pos=t]
\centering
\includegraphics[scale=0.8]{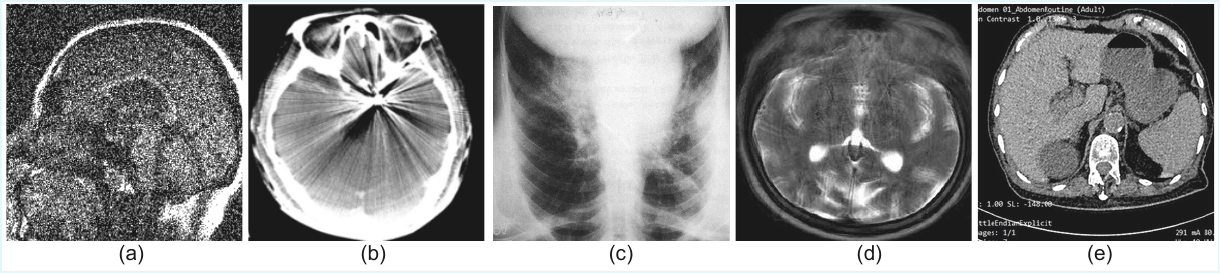}
\caption{Illustration of different factors affecting image quality in medical imaging. (a) Noisy brain MR image showing the degradation of fine structural details due to random noise, which can obscure subtle pathological changes \cite{khan2019mri}; (b) Metal artifact caused by dental fillings or implants, producing bright streaks that interfere with adjacent tissue visualization \cite{hung2014artifact}; (c)  Respiration blur resulting from patient breathing motion, reducing spatial resolution and anatomical clarity, particularly in thoracic imaging \cite{kim1997two}; (d)  Motion artifact from voluntary or involuntary patient movement during acquisition, leading to ghosting or distortion of structures \cite{rauf2018improve}; (e) Low-Dose \ac{CT} abdominal image showing increased noise and reduced contrast, demonstrating the trade-off between radiation dose and diagnostic image quality \cite{zhang2024innovative}.} 
\label{figsnoise}
\end{figure}

Image quality is a fundamental determinant of diagnostic accuracy in medical imaging, as it directly influences clinicians’ ability to detect, characterize, and monitor anatomical structures and pathological conditions. High-quality images provide clear visualization of tissues, lesions, and vascular networks, enabling precise diagnosis and effective treatment planning, while poor image quality caused by noise, artifacts, motion blur, or low-dose acquisition can obscure key structures and lead to misinterpretation or missed findings \cite{clement2025ai, alkhodari2025enhancing}. Figure \ref{figsnoise}, collectively highlight the diverse sources of image degradation in medical imaging and their potential to compromise diagnostic accuracy. Understanding and mitigating such effects are essential for optimizing image acquisition protocols, ensuring that even under constrained conditions such as low-dose or portable imaging clinicians can obtain diagnostically reliable images. In the following, we present a general equation that expresses how noise affects a medical image. This is the fundamental noise model used in medical imaging. It assumes that every pixel intensity in the observed image is the sum of the actual signal and random noise introduced during acquisition, transmission, or reconstruction. The noise term $n(x,y)$ can follow different statistical distributions depending on the imaging modality for instance, Gaussian noise in \ac{MRI} and \ac{US}, or Poisson noise in \ac{CT} and nuclear medicine. The fundamental noise model in medical imaging is given by Equation \ref{eq:noise}:

\begin{equation}
    I(x, y) = S(x, y) + n(x, y)
    \label{eq:noise}
\end{equation}

Where, $I(x, y)$ the observed noisy image, $S(x, y)$ the true noise-free signal, and $n(x, y)$ the noise component added to the signal.  In \ac{PMI}, image quality is even more crucial because these systems are often used in point-of-care, emergency, or resource-limited settings where fast and accurate decisions are essential. Unlike stationary imaging systems, portable devices such as \ac{PUS}, \ac{PCT}, and \ac{PMRI} systems are designed for mobility, compactness, and ease of use, often at the expense of certain performance parameters. However, reduced size and power limitations typically lead to lower \ac{SNR}, limited spatial resolution, and higher susceptibility to artifacts, all of which can significantly impact diagnostic reliability \cite{kravchenko2025low, shaddock2022potential, rao2024point}. For instance, \ac{PCT} systems are indispensable in intensive care units or disaster zones, but their images may suffer from motion blur due to lack of patient immobilization and noise caused by lower radiation doses to ensure patient safety. In low-dose \ac{PCT} scanners, reducing radiation exposure further amplifies image noise and reduces spatial resolution, making it harder to detect subtle pathologies such as micro-fractures or early lung nodules. Low-dose SPECT reduces patient radiation exposure by lowering tracer activity or scan duration, but this leads to higher noise, lower signal-to-noise ratio, and reduced spatial resolution. These limitations produce blurred or grainy images, making lesion detection harder, though iterative reconstruction methods can partially improve diagnostic image quality \cite{goertz2024state, ritt2022recent}. \ac{MRI} offers multiple imaging modes, each tailored to highlight specific tissue characteristics and diagnostic needs. T1-weighted images provide excellent anatomical detail, showing fat as bright and fluid as dark, while T2-weighted images make fluids appear bright, ideal for detecting edema or inflammation. proton density imaging balances T1 and T2 contrasts, useful in musculoskeletal studies, whereas diffusion-weighted imaging detects water molecule motion, essential for early stroke and tumor evaluation. FLAIR sequences suppress fluid signals to enhance lesion visibility near cerebrospinal fluid, particularly in brain imaging. Gradient echo sequences are sensitive to magnetic susceptibility and useful for detecting hemorrhages or calcifications. Functional \ac{MRI} measures brain activity through blood oxygen level–dependent contrast, and magnetic resonance angiography visualizes blood vessels without contrast agents. Together, these modes allow \ac{MRI} to provide both detailed structural and functional insights across a wide range of clinical applications \cite{khan2019mri}. \ac{PMRI} units, which often operate at lower magnetic field strengths, produce images with lower contrast-to-noise ratio and more artifacts, complicating the visualization of fine brain or musculoskeletal details \cite{bossert2023novel}. In \ac{PUS} imaging, speckle noise remains a major limitation, giving images a grainy appearance that can obscure small lesions, vessels, or organ boundaries. Since \ac{PUS} devices often rely on simpler transducers and lower-frequency probes, image resolution and penetration depth may be compromised. Additionally, image interpretation is strongly operator-dependent, meaning that diagnostic accuracy can vary based on the skill of the clinician \cite{dong2021feature}. While portable imaging systems greatly enhance accessibility bringing diagnostic capabilities to bedside, rural areas, or field operationstheir image quality limitations such as noise, artifacts, blur, low resolution can affect diagnostic confidence. Therefore, modern research focuses on \ac{AI}-based noise reduction, advanced reconstruction algorithms, and hardware miniaturization to bridge the performance gap between portable and conventional imaging systems. Ensuring high image quality in portable modalities is vital to achieve reliable, timely, and safe medical decisions in settings where every second and every pixel matters. \ac{MIQA} methods are categorized into \ac{FR}, \ac{RR}, and \ac{NR} approaches. \ac{FR} metrics compare degraded images with high-quality references, \ac{RR} metrics use partial reference features, and \ac{NR} metrics assess image quality directly without reference data. To ensure diagnostic reliability, these frameworks integrate statistical indicators and gold-standard comparisons, supporting consistent and accurate image evaluation across diverse imaging modalities and acquisition conditions, including constrained or emergency medical environments \cite{athar2023degraded, ohashi2023applicability, varga2021analysis}

\subsection{Image quality assessment}

\Ac{IQA} combines subjective and objective methods. Subjective evaluation depends on radiologists’ judgment, while objective analysis uses measures. In the following, both approaches are discussed to ensure optimal diagnostic performance and reliable medical imaging outcomes.

\subsubsection{Subjective quality assessment}
Subjective \ac{IQA} is a fundamental component of medical imaging evaluation, particularly in portable and low-dose imaging systems, where technical constraints can significantly affect diagnostic performance. Unlike objective metrics such as \ac{SNR}, which quantify measurable aspects of image quality, subjective assessment focuses on human visual perception and diagnostic confidence, reflecting how radiologists interpret images in real-world clinical settings. In this approach, experienced observers visually inspect and score images based on specific criteria such as sharpness, contrast, noise level, artifact presence, lesion visibility, anatomical detail, and overall diagnostic acceptability to determine whether the image is suitable for accurate diagnosis or needs reacquisition \cite{rodrigues2024objective, sang2025no}. Several standardized protocols exist for subjective \ac{IQA}, including the likert scale or numerical rating scale \cite{marzuki2024likert}, visual grading analysis \cite{papini2025effect}, mean opinion score \cite{gao2022image}, and the European guidelines on quality criteria for diagnostic radiographic images, which provide structured anatomical visibility standards. These protocols allow radiologists to judge whether images from portable or low-dose systems meet minimum diagnostic criteria, despite limitations in resolution or contrast.
Ultimately, subjective \ac{IQA} bridges the gap between technical performance and clinical usability, translating physical image metrics into meaningful diagnostic feedback. In the context of \ac{PMI}, where operational flexibility often comes at the cost of image fidelity, this approach ensures that acquired images remain clinically acceptable, diagnostically reliable, and safe for patients, supporting accurate decision-making in diverse environments such as intensive care units, rural clinics, and disaster response scenarios.

\subsubsection{Objective IQA metrics}
Objective image quality metrics in medical imaging use both traditional and \ac{AI}-based approaches to evaluate diagnostic accuracy and system performance. Traditional \ac{IQA} relies on physics-based and mathematical evaluations, focusing on parameters such as image sharpness, tissue boundaries, contrast visibility, and background noise. These metrics are useful for comparing imaging protocols and ensuring consistency across modalities like \ac{CT}, \ac{MRI}, and \ac{US}, but they don't fully capture radiologists' perceptions, especially in complex or noisy environments. For example, \ac{PSNR}, which quantifies reconstruction fidelity by measuring pixel-wise error; \Ac{SSIM}, which evaluates structural similarity based on luminance, contrast, and structure comparisons; \Ac{FSIM}, which emphasizes feature similarity using phase congruency and gradient information; \Ac{VIF}, which estimates the amount of visual information preserved relative to a reference image; \Ac{MAE}, measures the average absolute difference between original and distorted images, indicating overall reconstruction accuracy without emphasizing large errors disproportionately; \Ac{NCC}, which measures statistical correlation between images; and BRISQUE, a no-reference metric that assesses perceptual quality based on natural scene statistics without requiring an original image \cite{zhu2020deep, zhou2019image, deng2024portable}. In contrast, \ac{AI}-based quality assessment employs \ac{ML} algorithms that learn patterns from large datasets of medical images evaluated by experts. These models predict perceived diagnostic quality, bridging the gap between technical precision and human perception. \ac{AI}-driven approaches are increasingly used in clinical practice to optimize imaging parameters, detect suboptimal scans, and support real-time quality assurance.

\subsection{PMID benchmark datasets and IQA protocols }
 
Researchers used several datasets to test their proposed models, including well-known public datasets, local datasets, and data from \ac{PMI}. Including data from \ac{PMI}s helped evaluate the models' performance in real and varied situations. This made the models better at working with medical images taken by these portable devices. Using different data sources improved the accuracy and reliability of the results and made the models more suitable for practical use in mobile healthcare settings. Table \ref{tab:datasets} presents a summary of these datasets, highlighting the source, characteristics, number of patients,  number of images,  image size,  and image format.

\section{PMI applications}
This section examines the use of \ac{PMI} across diverse domains, including \ac{FPGA} design, disease detection, image denoising, segmentation, and other related applications. Table~\ref{tab:applications} compiles a summary of these applications, outlining their performance as well as their respective strengths and limitations. It is worth noting that these applications were conducted in isolated, offline environments, and the \ac{PMI} was deployed in geographically independent locations away from hospitals and healthcare facilities.

\begin{table}[ht]
\centering
\tiny
\scriptsize
\caption{List of publicly available datasets used for \ac{PMI} applications}
\begin{tabular}{m{0.5cm} m{1cm} m{8cm} m{0.5cm} m{0.5cm} m{1cm} m{1cm} m{0.7cm} m{0.6cm}}
\label{tab:datasets} \\
\hline

IM & Dataset &Characteristics & NoP & NoI & IS & IF & UB& Source \\
\hline

\ac{US} &BUSI  & - The dataset consists of medical breast cancer images obtained through \ac{US} scans. It is divided into three categories: normal, benign, and malignant images.
 & 600 & 780 & 500x500 & PNG & \cite{jimenez2024branet} & Link \tablefootnote{https://www.sciencedirect.com/science/article/pii/S2352340919312181}\\

 & & - This dataset contains BUSI labeled benign or malignant, used for \ac{AI} research in medical image classification and diagnosis. & - & 250 & - & BMP & \cite{jimenez2024branet} & Link \tablefootnote{https://data.mendeley.com/datasets/wmy84gzngw/1}\\

 & & - Provides comprehensive breast \ac{US} cases, clear visuals, interventional guidance, and educational tools for radiology professionals. & - & 306 & 377$\times$396 & JPEG & \cite{jimenez2024branet} & Link\tablefootnote{https://www.auntminnie.com/clinical-news/article/15565901/breast-ultrasound-and-us-guided-interventional-techniques-a-multimedia-teaching-file}\\

\ac{US} & CHUD  & - The dataset contains \ac{US} images from knuckles and heart regions, captured for training, validation, and testing purposes with varying sizes and quality.
 & - & 500 & 180$\times$180 \newline 321$\times$481 \newline 440$\times$380 & - & \cite{dong2021feature} & Link\tablefootnote{https://clarius.com/fr/}\\

\ac{US} & RSCD  & - The dataset contains diverse abdominal \ac{US} images of liver, kidney, and gallbladder, supporting accurate classification using \ac{DL} on mobile devices. & 25 & 38,065 & 800$\times$600 & DSC & \cite{lee2021reverse} & Link\tablefootnote{https://www.kdca.go.kr/index.es?sid=a3}\\

\ac{CT} & GCL  & - The dataset includes chest x-rays of COVID-19 patients, those with similar lung diseases, and patients without these specific conditions, all labeled by experts.
 & - & 6302 & 5600x4700 & JPEG & \cite{vidal2021multi} & Link\tablefootnote{https://arxiv.org/pdf/2006.11988}\\

\ac{CT} & COVID-19 pCXR  & - The dataset includes diverse frontal chest x-rays of COVID-19 patients, varying in resolution and view, collected from multiple public sources. & 197 & 673 & - & JPEG/PNG & \cite{kikkisetti2020deep} & Link\tablefootnote{https://arxiv.org/abs/2003.11597}\\

\ac{CT} & VinDr-Mammo  & - The VinDr-Mammo dataset offers large-scale, annotated full-field digital mammograms with BI-RADS assessments for advanced breast cancer detection research.
 & 5,000 & 20,000 & - & DICOM & \cite{jimenez2024branet} & Link\tablefootnote{https://physionet.org/content/vindr-mammo/1.0.0/}\\

\ac{CT} & INBreast  & - The INbreast database contains full-field digital mammograms with expert annotations, diverse lesions, and accurate contours for breast cancer research.
 & 115 & 410 & - & PGM & \cite{jimenez2024branet} & Link\tablefootnote{https://biokeanos.com/source/INBreast}\\

\ac{CT} & mini-MIAS  & - The mini-MIAS database contains digitized mammograms with standardized image size and detailed abnormality annotations for use in breast cancer research.
 & 161 & 322 & 1024$\times$1024 & PGM & \cite{jimenez2024branet} & Link\tablefootnote{http://peipa.essex.ac.uk/info/mias.html}\\

 \ac{CT} & CBIS-DDSM  & - The database offers extensive mammogram images, including normal and cancer cases, to support breast cancer screening research and analysis. & 2620 & 10,480 & 43–50 mpp & TIFF/PGM & \cite{jimenez2024branet} & Link\tablefootnote{https://link.springer.com/chapter/10.1007/978-94-011-5318-8\_75} \\
\hline
\end{tabular}
\begin{flushleft}
\textbf{Abbreviations:} 
NoP: number of patients, BUSI: breast \ac{US} images dataset, NoI: number of images, IS: image size, IF: image format, CBIS-DDSM: curated breast imaging subset–digital database for screening mammography,  PGM: portable gray map, MIAS: mammographic image analysis society, mpp: microns per pixel, GCL: general COVID lung, CHUD: clarius handheld \ac{US} device, RSCD: reverse scan conversion dataset, UB: Used by, IM: image modality.
\end{flushleft}
\end{table}

\subsection{Disease detection and classification in PMI}
\subsubsection{Disease detection}

\textbf{(a) Detection in \ac{PMRI}:}
The study by \cite{cooley2021portable} introduces a portable, low-field (80 mT) brain \ac{MRI} scanner using a 122-kg Halbach permanent magnet with a built-in gradient, requiring no cryogenics and only $\sim 800 W$ power. It produces T1, T2, and PD-weighted images at 2.2$\times$1.3$\times$ 6.8 mm$^3$ resolution within $\sim10–20$ minutes and low acoustic noise. Despite lower \ac{SNR} and some geometric distortions compared to high-field \ac{MRI}, the system enables bedside imaging. Importantly, it can potentially detect and monitor neurological diseases such as stroke, hemorrhage, hydrocephalus, tumors, and traumatic brain injury, improving access to neuroimaging in critical and resource-limited settings. In addition, the study by \cite{sheth2022bedside} evaluated a 0.064 T portable \ac{MRI} for bedside detection of intracranial midline shift (MLS) in 102 stroke and hemorrhage patients in the Yale NICU. Importantly, pMRI-based MLS measurements predicted poor discharge outcomes (adjusted OR 7.98). The scanner enabled safe, in-room imaging without patient transport. 

\textbf{(b) Detection in \ac{PCT}:}
In addition, this study \cite{david2023comparison} aimed to evaluate the diagnostic accuracy of an \ac{AI} system for interpreting portable chest x-rays in diagnosing COVID-19, comparing it with human readers. A total of 94 patients, including 65 with confirmed COVID-19, were included, with high-resolution chest \ac{CT} serving as the gold standard when available. Surprisingly, the \ac{AI} system demonstrated high sensitivity, particularly in detecting pleural effusion and consolidation, with moderate specificity for both conditions. The \ac{AI}'s performance was comparable to human radiologists, with significant agreement for pleural effusion and consolidation, but less accurate for other pathologies such as mediastinum widening or lung collapse.

\subsubsection{Classification}
Classification is widely used in medicine through pretrained models that learn patterns in medical images. These models speed up diagnosis and reduce human errors. Portable medical devices, allow doctors to use these models for quick and accurate diagnoses in various locations, improving access to healthcare. 

\textbf{(a) Classification in \ac{PCT}:}
For example, Wang et al. \cite{wang2024deep} developed a \ac{DL}-based \ac{CAD} system to accurately localize nasogastric tubes in portable supine chest x-rays in emergency and intensive care settings. The study used a dataset of 7,378 chest x-rays from two hospitals, with pixel-level annotations for nasogastric tubes localization and image-level labels for nasogastric tubes presence and malposition. The \ac{CAD} system utilized DeepLabv3+ for segmentation and DenseNet121 for classification. Despite high reported performance, the study is limited by severe class imbalance, very few malposition cases in testing datasets, potential selection bias, and restricted external generalizability. Suboptimal diaphragm segmentation and lack of prospective clinical validation further constrain real-world applicability. Moreover, the paper \cite{ieracitano2022fuzzy} presents a novel fuzzy-enhanced \ac{DL} approach for early detection of Covid-19 pneumonia from portable chest x-ray images. The authors developed a hybrid model, CovNNet, which combines \ac{CNN} with fuzzy edge detection techniques to improve classification performance. By integrating chest x-ray images and fuzzy-enhanced features, the model effectively distinguishes between Covid-19 and non-Covid-19 interstitial pneumonias. The proposed approach outperforming traditional methods. The study also emphasizes the importance of explainable \ac{AI} techniques, utilizing saliency maps to highlight relevant regions in chest x-ray images, enhancing the interpretability of the model's predictions. This research demonstrates the potential of combining fuzzy logic and \ac{DL} in portable diagnostic tools for rapid and accurate Covid-19 detection, contributing to better triaging in acute clinical settings.

\textbf{(b) Classification in \ac{PUS}:}
Similarly, the study \cite{zemi2024assessing} aimed to explore the feasibility of integrating \ac{AI} algorithms into \ac{POCUS} devices for breast cancer detection, achieving a performance benchmark of at least 15 FPS. The research utilized five \ac{AI} models (FasterRCNN+MobileNetV3, FasterRCNN+ResNet50, RetinaNet+ResNet50, SSD300+VGG16, and SSDLite320+ MobileNetV3), which were pretrained on natural image datasets and fine-tuned using a set of gelatin-based breast phantom images. Notably, these images contained anechoic and hyperechoic lesions, mimicking real tissue. The Clarius L15 \ac{US} probe, connected to a tablet and laptop, streamed scanning videos in real-time, with 200 timing trials conducted for each model. Ultimately, the SSDLite320+MobileNetV3 model performed best, making it suitable for real-time use in \ac{POCUS}.

Regarding pretrained models, they are effective in saving time and computational resources. However, they may face challenges such as overfitting when fine-tuned on small or specific datasets, which can impact the accuracy of predictions in certain cases.

\color{black}

\subsection{Image denoising in PMI}
Image denoising and enhancement play a crucial role in improving the quality and accuracy of \ac{PMI}, where the compact nature of devices often leads to lower-resolution images and increased noise. These techniques help refine images by reducing artifacts, improving contrast, and enhancing the clarity of important details, such as small lesions, fractures, or vascular structures. In portable \ac{US}, x-ray, and \ac{MRI} devices, denoising algorithms remove interference from environmental factors or patient movement, while enhancement methods ensure better visualization of tissues and organs. This improves diagnostic confidence, enabling healthcare professionals to make more accurate decisions even in challenging settings, such as emergency rooms or remote locations with limited resources \cite{dong2021feature, boucherit2025reinforced}. The work in \cite{zhu2024advancing} introduces ImT-MRD, a universal complex-valued imaging transformer designed to denoise low-field MRI (0.3T–0.55T) and overcome intrinsic low \ac{SNR} limitations. Trained on 8,583 diverse 3T scans, the model demonstrated strong generalizability across organs, sequences, and field strengths.

\subsection{Segmentation and prediction in PMI}
Segmentation-based techniques in \ac{PMI} are crucial for isolating and analyzing specific region of interest, such as organs, tissues, or abnormalities like tumors or fractures, from the surrounding background. In portable \ac{US}, \ac{CT}, and \ac{MRI} devices, segmentation helps to automatically identify and delineate structures such as blood vessels, tumors, or heart chambers, reducing the need for manual intervention and speeding up the diagnostic process. 

\textbf{(a) Segmentation in \ac{PMRI}:}
Also, in \cite{bian2024quantitative}, the authors employed is image segmentation, where the SCUNet \ac{DL} model is utilized to segment ischemic lesions in \ac{MRI} images. This segmentation allows for precise identification and quantification of lesion volumes and facilitates the calculation of stroke assessment scores like ASPECTS. By enhancing low-field strength \ac{MRI} (LF-MRI) through super-resolution to generate SynthMRI images, the study significantly improves lesion detection accuracy, particularly during critical post-stroke phases, when compared to conventional LF-\ac{MRI} scans. In addition, this study \cite{sheth2020first} introduces a groundbreaking application of portable, low-field \ac{MRI} technology for bedside assessment of acute brain injury. By using \ac{AI}-driven analysis, specifically a \ac{DL} model based on the U-Net neural network, the research demonstrates the feasibility of evaluating mass effect from 3D \ac{MRI} images without the need for traditional high-field \ac{MRI} systems. The study involved 66 ICU patients who underwent portable 64 mT \ac{MRI} scans at the bedside, with \ac{AI} interpreting midline shift as a measure of mass effect. The results revealed strong correlation between human and \ac{AI} assessments, suggesting that portable \ac{MRI} could replace traditional methods of assessing mass effect in emergency settings.

\textbf{(b) Segmentation in \ac{PCT}:}
Similarly, this research \cite{moris2024multi} introduces a novel \ac{DL} methodology designed to simultaneously localize hemidiaphragms' landmarks and segment lung boundaries in portable chest x-ray images of COVID-19 patients. By employing a multi-task learning approach, the study utilizes a heatmap regression technique for landmark detection and a U-Net architecture for lung segmentation, offering significant improvements in diagnostic assistance despite the lower quality of \ac{PCT} images. The methodology is validated through extensive testing, showing promising results for hemidiaphragm localization and lung segmentation. This innovative approach can help clinicians rapidly assess diaphragmatic function and monitor lung conditions in COVID-19 patients, contributing to more effective management during the pandemic and beyond. In a similar vein, the paper \cite{rogelio2022object} by Rogelio et al. presents a \ac{DL}-based approach for detecting and segmenting parts of a \ac{PCT} source model, such as the body, handle, and aperture, using the Deeplabv3 architecture. The study demonstrates that the Deeplabv3 deep neural network can effectively segment the body with high precision, but faces challenges in segmenting the handle and aperture due to their smaller size and lower resolution. Despite this, the authors show the potential of using atrous convolution for improving segmentation accuracy. Their results indicate that the model is successful, though further improvements are needed. The findings provide valuable insights for enhancing automatic x-ray film alignment in robot-assisted operations and suggest avenues for future research, including the integration of depth images for more accurate segmentation.

\textbf{(c) Segmentation in \ac{PUS}:}
For example, this study \cite{vaze2020low} focuses on addressing the limitations of \ac{CNN}s in real-time \ac{US} segmentation for mobile deployment. \ac{CNN}s, though highly effective for image analysis, have large memory footprints and computational demands that hinder their use in mobile or clinical settings. To overcome these challenges, the authors propose a series of techniques to develop memory-efficient \ac{CNN}s that maintain high performance. These include reducing the number of feature channels in \ac{CNN} layers, utilizing depth-wise separable convolutions for efficient computation, and implementing knowledge distillation to improve accuracy while preserving computational efficiency. In contrast, the article \cite{fournelle2021portable} presents the development of a portable \ac{US} research system (MoUsE) designed for automated bladder monitoring, leveraging \ac{ML}-based segmentation to assist in detecting postoperative urinary retention (POUR) and other bladder-related complications. The system integrates a compact 32-element phased array transducer and innovative beamforming algorithms, offering real-time image reconstruction through a GPU-based approach. A key focus of the study was the development of a U-Net-based \ac{ML} model for bladder segmentation, which was trained with a limited dataset of 253 images, achieving promising results for bladder volume estimation. However, further refinement in segmentation accuracy and data handling for mobile applications is needed. 
The research conducted by Lin et al.  \cite{lin2023handheld} presents a novel dual-modality \ac{US} and photoacoustic (PA) navigation system aimed at improving the visibility and tracking accuracy of puncture needles during interventional procedures. The study addresses the common challenges of acoustic interferences in \ac{US}, which often lead to blurred or invisible needle tips. By incorporating internal illumination to highlight the needle tip in photoacoustic imaging, the system enhances the needle’s visibility. A \ac{DL}-based segmentation model, AG U-Net++, is employed to refine the needle detection, removing artifacts and improving the precision of needle tracking. The combination of \ac{US} and photoacoustic imaging offers complementary contrasts, allowing for better localization and registration of both the needle shaft and tip. The system demonstrates submillimeter target registration accuracy and has been successfully validated through phantom, ex vivo, and in vivo experiments, suggesting its potential for clinical applications in precise needle-guided interventions such as biopsies and radiofrequency ablation.

Segmentation in \ac{PMI} is valuable for precise analysis, but it may struggle with low-quality images or variability in patient organs. Over-reliance on automated segmentation could also lead to missed details that a healthcare professional might otherwise identify.

\textbf{(d) Prediction in \ac{PMI}:}
For prediction, its involves using a set of models to forecast medical conditions or outcomes from images captured by portable devices. These predictions help in early diagnosis, aiding healthcare providers in making informed decisions quickly, even in remote settings. For example, the paper \cite{hwang2020smartphone} presents an innovative approach for diagnosing diabetic macular edema (DME) using a smartphone-based \ac{AI} model. The model, built on the MobileNet architecture and optimized for mobile and embedded systems, analyzes optical coherence tomography (OCT) images to predict the likelihood of DME in patients. It offers diagnostic suggestions and medical recommendations for those at risk, aiding in early detection. The model, trained using a \ac{CNN}, shows high diagnostic performance comparable to more complex \ac{AI} systems. Additionally, the system is embedded in a mobile application that operates offline, making it accessible in areas with limited healthcare resources.

Prediction in \ac{PMI} is a powerful tool, but it may face challenges like model accuracy when dealing with low-quality images or incomplete data. Additionally, over-reliance on automated predictions might diminish the role of human assessment in some cases.

\color{black}

\begin{table}[!t]
\centering
\tiny
\scriptsize
\caption{Summary of the performance and limitations of specific \ac{PMI} applications. In cases where multiple
results are conducted, only the best performance is reported.}
\begin{tabular}{m{0.3cm} m{0.3cm} m{1.5cm} m{8cm} m{2cm} m{3.2cm} m{0.3cm}}
\label{tab:applications} \\
\hline

Ref. & Year & Application &Description & BP & Limitations & PLA  \\
\hline

\cite{wang2024portable}&2024& Segmentation&\ac{FPGA}-based handheld {US} system using synthetic aperture imaging, with auto delay calculation and segmented apodization.&Frame rate: 22 fps&Lower SNR and contrast \newline Low {FPGA} power: $\sim530mW$&No\\
\hline

\cite{cooley2021portable}&2021& Detection&Portable, low-field (80 mT) head-only {MRI} using 122-kg Halbach permanent magnet with built-in gradient, enabling point-of-care neuroimaging&Power: $\sim 800 W$&Image distortions from non-linear encoding fields.& \href{https://github.com/czcooley/portable-MRI}{Yes}\\
\hline

\cite{sheth2022bedside}&2022& Detection &Bedside low-field portable {MRI} (0.064 T) evaluated for detecting intracranial midline shift in ischemic stroke and hemorrhage patients.&Sen:0.93 \newline Spe:0.96&Some exams limited by motion artifacts&No\\
\hline

\cite{zhu2024advancing}&2024& Denoising&Universal complex-valued imaging transformer (ImT-MRD) denoises low-field {MRI}, enabling faster scans, improved image quality, and broad clinical generalization.&PSNR 36.8dB \newline SSIM 0.95&Motion artifacts and rare sequences still challenging&\href{https://github.com/zherenz/ImT-MRD.}{Yes}\\
\hline

\cite{vaze2020low}&2020& Segmentation&Lightweight {CNN}s using thin architectures, separable convolutions, and knowledge distillation for real-time {US} segmentation on CPUs/mobile devices.&30 fps \newline Dice: 83.7\%&Limited to 2D {US} nerve segmentation&\href{ https://github.com/sagarvaze96/lightweight_unet}{Yes}\\
\hline

\cite{fournelle2021portable}&2021& Segmentation&Portable {US} research system (MoUsE) with 32-element phased array and {CNN}-based segmentation for automated postoperative bladder monitoring.&IoU:0.75&Small dataset (20 test images, 253 training)&No\\
\hline

\cite{moris2024multi}&2024& Segmentation&Multi-task {DL} method for hemidiaphragm localization and lung segmentation in portable chest x-rays of COVID-19 patients.&Acc: 82.31\% \newline Dice: 0.9688\%&Dataset limited to 673 images&No\\
\hline

\cite{rogelio2022object}&2022&Segmentation \newline Detection &Implemented Deeplabv3 {CNN} for object detection and segmentation of {PCT} source model parts (body, handle, aperture).&IoU: 76.42\% \newline mAP: 39.47\%. &Weak detection of small/low-resolution parts (handle, aperture)&No\\
\hline

\cite{lin2023handheld}&2023& Segmentation &Dual-modality {US} imaging with internal optical fiber illumination and AG U-Net++ segmentation for precise handheld needle tip tracking.&IoU: 0.8361 \newline Dice: 0.8785\%&Dataset relatively small&No\\
\hline

\cite{lucas2023multi}&2023&Segmentation &LowGAN architecture uses a generative adversarial network to translate low-field {MRI} images (64mT) to high-field quality images for tissue segmentation & p < 0.001& Underperformance in some regions &No \\\hline
\cite{islam2023improving}&2023 &Segmentation& LoHiResGAN for enhancing low-field (64mT) {MRI} scans, generating high-quality synthetic 3T {MRI} images& Dice > 0.9& Challenges in small structures& \href{ https://github.com/khtohidulislam/LoHiResGAN}{Yes} \\\hline
\cite{lommen2021accuracy}&2021 &Classification &Comparison of handheld vs wall-mounted X-ray &p < 0.001 &Minor scattering in handheld device &No \\\hline
\cite{li2023automatic}&2023&Detection &Portable freehand 3D {US} imaging system with a U-Net segmentation algorithm for detecting carotid atherosclerosis. &Sen:71.00\%,\newline Spe:85.00\%,\newline Acc:80.00\% & High sensitivity and requires manual correction& No \\\hline
\cite{moris2021data}&2021 &Classification & CycleGAN-based data augmentation for generating synthetic portable chest X-ray images, improving COVID-19 screening.& Acc:99.00\%, \newline Recall: 99.00\%, \newline F1:99.00\%&Challenges with low-quality images &No \\\hline
\cite{cores2022few}&2022 & Classification&Few-shot learning method with a lung-aware region proposal network for COVID-19 detection &Acc:79.10\% & Sensitive to dataset imbalance&No \\
\hline
\cite{cho2023lightweight}& 2023 & Classification \newline Segmentation & Network optimized for {PUS} systems using a {CNN}-based multitask architecture& Dice: 0.954&Accuracy depends on shape coefficients &No \\
\hline
\cite{kessler2024development}&2024 & Detection \newline Classification& A {DL} algorithm was developed to detect lung consolidation using handheld {US} devices. The algorithm employs VGG-like CNN to classify each frame and utilizes a MobileNet V3 CNN to detect  consolidation features.& Sen:88.00\% \newline Spe:89.00\%& Small Training Set& No\\
\hline
\cite{bian2025quantitative}&2025 & Detection & A {DL}-based Swin-Conv-UNet was used to generate SynthMRI images from LF-MRI, improving the detection and quantification of ischemic lesions.&Sen:89.00\% \newline Spe: 91.30\% & Single LF-MRI scanner type& \href{https://
github.com/longw010/Low_Field_Enhancement}{Yes}\\
\hline

\end{tabular}
\begin{flushleft}
\textbf{Abbreviations:} 
 Best performance (BP); Project link availability (PLA). 
\end{flushleft}
\end{table}

\section{Research challenges and future directions}

\subsection{Research challenges}

\medskip
\textbf{Security compliance of \ac{PMI}:} Security compliance of \ac{PMI} systems is a critical aspect of ensuring patient safety, data integrity, and regulatory adherence in modern connected healthcare environments. These devices, often integrated with \ac{IoT} networks and cloud platforms, handle highly sensitive medical images that must comply with international security standards such as HIPAA, GDPR, ISO/IEC 27001, and FDA cybersecurity guidelines \cite{harris2024cybersecurity, montell2025roadmap}. Compliance requires implementing strong encryption for data transmission and storage, robust authentication and access control mechanisms, and regular vulnerability assessments to prevent unauthorized access or data breaches. Additionally, maintaining audit trails, secure firmware updates, and role-based user management ensures accountability and resilience against cyber threats. With the growing adoption of remote diagnostics and mobile imaging, ensuring end-to-end security from device to cloud is essential. Adherence to regulatory frameworks not only protects patient confidentiality but also fosters trust in digital healthcare ecosystems, enabling safe and reliable deployment of portable imaging technologies worldwide \cite{wang2020security, tettey2024review}.

\medskip
\textbf{Medical device miniaturization:} Miniaturizing medical imaging devices to make them portable presents a fundamental challenge: balancing device size with image quality. High-resolution imaging relies on powerful sensors and intensive data processing, which typically increase the device’s size and energy consumption. Reducing the size often compromises sensor capability and image accuracy, directly impacting diagnostic quality. Therefore, effective miniaturization demands advanced technologies that strike a balance between compactness and maintaining image clarity. The goal is to ensure precise diagnostics while enabling ease of transport and use, especially in areas lacking advanced medical infrastructure. This approach paves the way for new possibilities in portable and field healthcare delivery \cite{wang2019high, yang2025miniaturized}.

\medskip
\textbf{Dynamic digital radiography:} Dynamic digital radiography represents a significant advancement in diagnostic imaging, relying on pulsed x-rays to capture a sequential series of high-resolution images of the targeted anatomical area. Initially used for chest imaging in a standing position, this technique has recently expanded due to the development of portable devices that can be positioned beside the patient’s bed, enabling easier examination of immobile patients. This equipment provides precise anatomical and functional information about the movement of the lungs, pleura, rib cage, and trachea, with specialized software allowing further analysis to extract quantitative parameters such as diaphragm motion, projected lung area, rate of area change, and color maps reflecting respiration and lung perfusion. However, dynamic radiography faces significant challenges, including achieving a delicate balance between image quality and minimizing patient radiation exposure, as well as developing portable devices that effectively suit bedside environments. Managing the large volume of generated data requires accurate extraction of indicators and reliable interpretation of results. Additionally, ensuring the technique’s portability and versatility across diverse clinical settings while maintaining affordable costs is crucial. Despite these challenges, dynamic digital radiography remains a promising technology poised to revolutionize diagnostic imaging and bedside monitoring, enhancing physicians’ ability to provide better patient care \cite{yamasaki2020novel, cellina2022artificial}.

\medskip
\textbf{Limited computational resources:} 
\ac{PMI} devices face a critical challenge limited computational resources. Unlike stationary, hospital-grade systems equipped with powerful processors and extensive memory, portable devices must balance performance with size, power consumption, and heat dissipation constraints. This limitation impacts image acquisition, processing speed, and the ability to run advanced algorithms like \ac{AI}-based diagnostics or real-time 3D reconstructions. Consequently, developers must optimize software for efficiency, often sacrificing some image quality or processing complexity. Battery life is another bottleneck high computational demand drains power quickly, reducing device usability in remote or emergency settings. Limited storage also restricts data retention and transmission capabilities. Overcoming these constraints requires innovative hardware-software co-design, such as lightweight neural networks, edge computing strategies, and specialized low-power chips. Addressing limited computational resources is essential to making portable imaging devices truly effective and accessible, particularly in resource-poor environments where their impact can be life-saving \cite{ranger2024portable}, \cite{ fournelle2021portable}.

\medskip
\textbf{Battery life and power management:} Battery life and power management are critical hurdles for \ac{PMI} devices. These devices must deliver high-quality imaging while operating within strict energy constraints, as bulky or heavy batteries undermine portability. High computational demands for image processing and real-time analysis rapidly drain power, limiting continuous use and field deployment. Efficient power management strategies such as adaptive processing that lowers computational load when possible, low-power hardware components, and energy-aware software algorithms are essential to extend operational time. Innovations like dynamic voltage scaling and duty cycling help balance performance with power consumption. Additionally, integrating energy-efficient wireless communication reduces power spent on data transmission. Optimizing battery technology itself, including fast charging and higher energy density options, further supports longer use in remote or emergency settings where recharging opportunities are scarce. Without robust power management, the practical utility of \ac{PMI} devices remains compromised, especially in low-resource environments where they are most needed \cite{daining2024application}, \cite{noohi2024design}.

\medskip
\textbf{Inappropriateness of metrics:} The \ac{PMI} environment faces unique challenges that make many traditional image quality metrics unsuitable for accurate performance evaluation. Classical quantitative indicators such as \ac{PSNR}, \ac{SSIM}, and \ac{MSE} were designed for controlled laboratory settings with ideal imaging conditions and do not account for variable field factors such as vibration, uneven lighting, high noise levels caused by motion, or power limitations. Moreover, these metrics primarily emphasize visual aspects of the image without considering its clinical diagnostic value, such as the clarity of anatomical structures or the precision in detecting pathological lesions. Additionally, the absence of high-quality reference images in many portable imaging scenarios renders the application of full-reference metrics impractical. Therefore, image assessment in \ac{PMI} environments requires intelligent and adaptive metrics that consider clinical context, device characteristics, and acquisition conditions to ensure truly diagnostic image quality that transcends purely technical standards.

\medskip
\textbf{Ethical and regulatory:} Ethical and legal aspects represent some of the most overlooked challenges in the field of \ac{AI}-driven \ac{PMI}. The collection of medical images in field environments involves significant risks related to data privacy and patient confidentiality, as images may be transmitted through unsecured networks or stored on cloud servers lacking adequate protection. Moreover, the algorithmic bias inherent in \ac{AI} models trained on unbalanced datasets can lead to discriminatory performance or reduced accuracy for certain population groups. In addition, the legal and regulatory framework governing the use and exchange of portable medical data remains unclear in many countries, raising concerns about liability in cases of diagnostic errors or data breaches. Therefore, the advancement of \ac{PMI} technologies requires the establishment of robust ethical and legal governance frameworks that ensure transparency, safeguard patient privacy, and promote fairness and accountability in intelligent diagnostic systems.

\medskip
\textbf{Technical challenges:} The \ac{PMI} environment faces multiple technical challenges that significantly affect image quality and diagnostic reliability. Limited computational power and energy constraints in portable devices hinder the real-time deployment of deep \ac{AI} models, emphasizing the need for lightweight and optimized algorithms. Moreover, unstable acquisition conditions such as vibration, poor lighting, and electromagnetic interference introduce noise and visual artifacts that are difficult to correct using traditional methods. Field data management also poses a challenge, as transmitting large medical images over wireless networks may lead to partial data loss or delays in cloud processing. In addition, inconsistent sensor calibration and varying device performance reduce uniformity and comparability across different imaging systems. Addressing these challenges requires the development of adaptive and intelligent systems that balance computational efficiency with environmental stability, ensuring accurate and reliable imaging across diverse field and clinical scenarios.

\subsection{Future directions}
With the continuous advancement of \ac{PMI}, future investigations must concentrate on addressing existing technological limitations while aligning innovations with clinical requirements. In the following, we highlight potential directions to enhance diagnostic reliability, accessibility, and adaptability of \ac{PMI} across diverse healthcare environments.

\medskip
\textbf{5G/6G Connectivity, and UAV Integration:} Fast, reliable wireless connectivity is transforming \ac{PMI} by enabling real-time data transfer and remote consultation. With 5G and emerging 6G networks \cite{kheddar2022efficient}, portable imaging devices can instantly stream high-resolution data to cloud platforms, where \ac{AI} algorithms perform automated analysis or specialists provide remote interpretation. This capability is particularly transformative for rural, disaster-stricken, or underserved regions lacking onsite radiologists. For example, a paramedic capturing \ac{US} images in the field can transmit them live to a hospital expert, accelerating diagnosis and intervention \cite{abolade2022miniaturized}. Beyond ground-based deployment, unmanned aerial vehicles (UAVs) equipped with lightweight thermal or optical imaging systems can capture high-resolution images of small or hard-to-reach targets in emergency or remote medical scenarios \cite{kheddar2025recent}. UAV-assisted medical imaging supports rapid triage, search-and-rescue operations, epidemic monitoring, and delivery of portable diagnostic kits. Combined with ultra-low latency, high reliability, person re-identification algorithms \cite{chouchane2024multilinear}, and cloud integration, these technologies position future portable devices as intelligent nodes within a fully connected healthcare ecosystem, bridging geographical, logistical, and expertise gaps in diagnostic imaging.

\medskip
\textbf{Extreme portability for low-resource settings:}  \ac{PMI} devices designed for extreme portability must withstand harsh environments while delivering reliable diagnostics. These rugged devices are vital in disaster zones, battlefields, and remote clinics where infrastructure is minimal. Future designs focus on durability shock-proof, dust-resistant, waterproof casings plus easy, rapid deployment with minimal calibration. Battery independence, simple user interfaces, and minimal maintenance are critical features. Self-calibrating sensors and automated quality checks ensure consistent performance even in non-ideal conditions. These devices empower healthcare workers in underserved areas to perform critical imaging tasks without relying on bulky, fragile machines or specialized operators. By providing accurate imaging on the move, they improve triage, treatment planning, and monitoring in emergencies and routine care, significantly impacting global health equity. The design philosophy prioritizes “plug and diagnose” functionality, making the technology accessible and practical for frontline responders and community health workers alike \cite{zhang2022smart}.

\medskip
\textbf{Cybersecurity and privacy enhancements:} As \ac{PMI} devices increasingly connect to networks and cloud platforms, cybersecurity becomes non-negotiable. The future focuses on robust security frameworks that protect sensitive patient data and device integrity against cyber threats. This includes on-device encryption of image data, secure boot processes, and intrusion detection systems tailored for medical devices \cite{gueriani2025robust}. Zero-trust architectures ensure that every data exchange is authenticated and authorized, minimizing risks from compromised endpoints. Furthermore, secure remote software updates keep devices resilient against emerging vulnerabilities.  Cybersecurity is not only a technical requirement but a trust imperative; patients and healthcare providers must be assured that portable imaging devices protect sensitive data against breaches and misuse through robust safeguards, including watermarking techniques \cite{kouadri2025robust}. Investment in security protocols will be critical to the widespread adoption of connected portable imaging solutions in clinical practice \cite{padmini2016improving, bibi2025cybersecurity}.

\medskip
\textbf{\Ac{LLM} for \ac{PMI}:} \ac{LLM}s hold immense promise for advancing \ac{PMI} by leveraging their capacity to learn intricate patterns from extensive and varied medical image datasets. Through extensive pre-training on diverse anatomical and pathological images, \ac{LLM}s develop a deep understanding of subtle features that traditional algorithms often miss. Fine-tuning these models on specific diagnostic tasks enhances their ability to accurately identify clinically relevant markers, improving detection rates for diseases and abnormalities in portable imaging contexts. Their strength lies in generalizing across different patient demographics, imaging conditions, and device types, which is crucial for robust performance in the field. Moreover, integrating \ac{LLM}s into portable devices can facilitate real-time, on-device analysis, reducing dependence on cloud computing and speeding up diagnostics. However, achieving this requires careful curation of high-quality datasets, optimizing model architectures for resource-constrained environments, and addressing stringent privacy and regulatory requirements around sensitive medical data to ensure patient confidentiality and compliance \cite{choi2023developing}.

\medskip
\textbf{Federated learning for \ac{PMI}:} Federated learning offers a promising future direction for \ac{PMI} by enabling decentralized model training across multiple devices without the need to share sensitive patient data. This technique allows medical imaging devices, such as portable \ac{US} or x-ray machines, to collaboratively improve diagnostic algorithms while preserving data privacy and security. In federated learning, each device trains the model on local data, and only model updates (rather than raw data) are shared with a central server for aggregation. This approach could lead to more personalized, real-time, and accurate diagnostic capabilities in resource-limited environments while ensuring compliance with data protection regulations. By leveraging the power of distributed learning, federated learning can help enhance the performance and adaptability of \ac{PMI} systems, making them more effective in diverse clinical settings \cite{himeur2023federated}.

\medskip
\textbf{Cloud computing:} For integrating \ac{PMI} with cloud computing involve enhancing real-time data processing through edge computing, where image analysis and initial processing occur locally on devices before transmitting to the cloud, minimizing latency and improving efficiency. Additionally, advancements in \ac{AI} and \ac{ML} will enable automated diagnostics, providing faster and more accurate readings of medical images. Interoperability between different imaging devices and cloud platforms must be improved, ensuring seamless integration and data sharing across systems. Security and privacy will remain a critical focus, with ongoing development of encryption and compliance measures to safeguard sensitive patient data. Finally, expanding the accessibility of cloud-based platforms to underserved regions will be essential, empowering remote healthcare delivery and facilitating global collaboration among medical professionals.

\medskip
\textbf{3D \ac{PMI}:} In the field of \ac{PMI}, 3D imaging is expected to make significant advancements in the coming years. With the use of advanced portable devices, doctors will be able to capture precise 3D images of organs and internal tissues in real-time. These portable devices will allow doctors to examine patients more quickly and accurately, aiding in the early detection of diseases or abnormalities. 3D imaging in these devices will enable doctors to see fine details of delicate structures such as blood vessels, nerve tissues, and joints, enhancing diagnostic and treatment precision. Moreover, this technology will make it easier to conduct exams in remote or under-equipped areas, providing faster and more comprehensive healthcare. A set of ideas were presented and embodied in many works \cite{huang2005development, tasinkiewicz20233d, ibrahim2018towards, chang2017portable, tasinkevych20193d}.

\medskip
\textbf{Multi-modal data:} Future research on multi-modal data for \ac{PMI} should focus on robust fusion of complementary signals such as thermal imaging and speech analysis. Thermal images can capture physiological markers related to inflammation, perfusion, or inhaling and exhaling patterns rhythm \cite{kheddar2025breathai}, while speech signals may reveal respiratory distress, fatigue, neurological impairment, or emotional state \cite{kheddar2024automatic,hamza2023machine}. Integrating these heterogeneous modalities through attention-based fusion networks, Transformer architectures, or cross-modal self-supervised learning can improve diagnostic accuracy and generalization. Lightweight and energy-efficient models are essential for real-time inference on edge devices with limited computational resources. Additionally, future work should address multi-modal dataset standardization, domain adaptation across populations, explainability of multimodal decisions, and privacy-preserving learning frameworks to ensure safe deployment in telemedicine and remote healthcare environments.

\medskip
\textbf{PMI compression:} Medical image compression is crucial for efficiently storing and transmitting medical images, particularly in portable devices used in healthcare settings. Compression techniques reduce the file size of images like \ac{CT} scans, and \ac{MRI}s while maintaining sufficient quality for accurate diagnosis. Lossless compression algorithms preserve every detail, ensuring no information is lost, while lossy compression can significantly reduce file size at the cost of some image quality, which may be acceptable for certain applications. Efficient compression enables faster data transmission, particularly in telemedicine, where remote diagnosis is becoming increasingly common. By using advanced algorithms, such as wavelet or transform coding, portable medical devices can manage large volumes of image data more effectively, improving accessibility, storage capacity, and operational efficiency in healthcare environments \cite{yee2017medical, mohammed2013comparative, beladgham2019medical}.

\medskip
\textbf{Standardized metrics:} In light of the technical and methodological challenges associated with assessing the quality of \ac{PMI}, there is an urgent need for a Clinically-Oriented \ac{MIQA} metrics, a unified and standardized approach for evaluating \ac{MIQA}. This framework should integrate traditional physical metrics (such as \ac{SSIM}, \ac{PSNR}, and \ac{SNR}) with actual clinical diagnostic indicators that reflect the image’s usability for safe and effective medical decision-making. The proposed future roadmap envisions a multi-level research strategy that includes: (1) developing standardized databases containing clinically annotated portable images collected under diverse environmental and operational conditions; (2) adopting Explainable \ac{AI} techniques to ensure that quality enhancement does not compromise diagnostically relevant information; (3) designing hybrid quality indices that connect visual features of the image with clinical decision outcomes; and (4) establishing unified evaluation protocols endorsed by regulatory and medical institutions. Building such a framework will enable the transformation of \ac{IQA} from a purely visual-technical process into an intelligent, clinically guided medical evaluation, thereby strengthening the reliability of portable imaging systems and ensuring their optimal use in field medicine and remote healthcare practices.

\section{Conclusion}
The field of \ac{PMI} is undergoing rapid transformation, driven by innovations in \ac{AI} and edge computing. These devices have redefined modern healthcare by providing mobility, accessibility, and rapid diagnostic capabilities across diverse environments from urban hospitals to remote and resource-limited settings. However, several persistent challenges continue to hinder their full diagnostic potential. Image quality remains a central concern, as environmental noise, hardware limitations, and operator variability frequently degrade accuracy and reliability. This review highlights the critical role of \ac{MIQA} frameworks in maintaining clinical validity within \ac{PMI} systems. Conventional full-reference and no-reference metrics remain valuable for benchmarking performance, yet they often overlook modality-specific distortions and clinically significant artifacts. The integration of \ac{AI} particularly \ac{DL}, \ac{TL}, and Transformers represents a paradigm shift in \ac{PMI} optimization. These intelligent models not only enhance image reconstruction, super-resolution, and denoising but also enable real-time analysis, adaptive calibration, and automated diagnostic support. Despite these advancements, the deployment of \ac{AI}-driven \ac{PMI} solutions faces notable challenges. Limitations in dataset diversity, computational efficiency, and model interpretability pose barriers to clinical translation. Furthermore, the lack of explainable \ac{AI} frameworks restricts user trust and regulatory acceptance. Moving forward, the development of lightweight architectures, modality-aware quality metrics, and transparent interpretability mechanisms will be vital for expanding \ac{PMI} applications in low-resource and point-of-care contexts. Ultimately, the convergence of \ac{AI} and \ac{PMI} signals a new era in intelligent healthcare imaging. By bridging technological innovation with clinical demands, future \ac{PMI} systems are poised to deliver equitable, efficient, and high-quality diagnostic services worldwide establishing themselves as indispensable tools in achieving global healthcare resilience and precision medicine.

\section*{Footnotes}


 

\subsection*{Conflict of Interest:}
The authors declare that they have no known competing interests or personal relationships that could have appeared to influence the work reported in this paper.

\subsection*{Consent to Publish Declaration:}
Not Applicable.
\subsection*{Consent to Participate Declaration:}
Not Applicable.

\printcredits

\bibliographystyle{elsarticle-num}
\bibliography{refs}

\end{document}